\documentclass[11pt]{article}

\def\bSig\mathbf{\Sigma}

\usepackage{epsfig}
\usepackage[round]{natbib}
\usepackage{amssymb}
\usepackage{amsmath}
\usepackage{multirow}
\usepackage{longtable}
\usepackage[figuresright]{rotating}
\usepackage[nolists]{endfloat}
\usepackage{endrotfloat}

\title{A Partitioning Deletion/Substitution/Addition Algorithm
for Creating Survival Risk  Groups}

\author{Karen Lostritto\footnote{Division of Biostatistics, Yale
   University School of Medicine, 60 College St., New Haven, 06519
\newline
\hspace*{0.18in}$\ddag$
Department of Statistical Science,  Cornell University, Ithaca, NY
\newline
\hspace*{0.18in}$\dag$
Departments of Neurological Surgery and Epidemiology and Biostatistics, University of California San Francisco, 505 Parnassus Ave, San Francisco, 94117},
  Robert L.\ Strawderman$\ddag$, and Annette M.\  Molinaro$^*$$\dag$ }

\date{}

\begin{document}

\maketitle

\begin{abstract}
  Accurately assessing a patient's risk of a given event is essential
  in making informed treatment decisions.  One approach is to stratify
  patients into two or more distinct risk groups with respect to a
  specific outcome using both clinical and demographic
  variables. Outcomes may be categorical or continuous in nature;
  important examples in cancer studies might include level of toxicity
  or time to recurrence. Recursive partitioning methods are ideal for
  building such risk groups.  Two such methods are Classification and
  Regression Trees (CART) and a more recent competitor known as the
  {\em partitioning Deletion/Substitution/Addition} (\emph{partDSA})
  algorithm, both which also utilize loss functions (e.g. squared error
  for a continuous outcome) as the basis for building, selecting and
  assessing predictors but differ in the manner by which regression
  trees are constructed.

  Recently, we have shown that \emph{partDSA} often outperforms
  CART in so-called ``full data'' (e.g., uncensored) settings.
  However, when confronted with censored outcome data,
  the loss functions used by both procedures must be modified.  There have
  been several attempts to adapt CART for right-censored data. This
  article describes two such extensions for \emph{partDSA} that make
  use of observed data (i.e. possibly censored) loss functions. These
  observed data loss functions, constructed using inverse probability of
  censoring weights, are consistent estimates of their uncensored
  counterparts provided that the corresponding censoring model is
  correctly specified.  The relative performance of these new methods
  is evaluated via simulation studies and illustrated through an
  analysis of clinical trial data on brain cancer patients.  The
  implementation of \emph{partDSA} for uncensored and right censored
  outcomes is publicly available in the \texttt{R} package,
  \texttt{partDSA}.\\
\end{abstract}


\section{Introduction}
\label{s:intro}

Clinicians carefully weigh a patient's prognosis when deciding on the
aggressiveness of treatment.  In determining a patient's prognosis,
clinicians may consider a patient's age, gender, clinical information
and, more recently, biological variables such as gene or protein
expression. Objective guidelines for predicting a patient's prognosis
(i.e., risk) from clinical and biological information can be obtained
from risk indices estimated from data collected on an independent
sample of patients with known covariates and outcome.  Frequently, the
outcome of interest is the time to occurrence of a specified event;
common examples in cancer include death and recurrence or progression
of disease. The use of time-to-event outcomes frequently results in
the presence of right-censored outcome data on several patients.

There are many statistical learning methods that might be used in
building predictors of risk for a given outcome using covariate
information.
An attractive class of methods for building clinically interpretable
risk indices is recursive partitioning methods.
The Classification and Regression Tree algorithm
\citep[CART;][]{Breiman1984} is perhaps the most
well-known recursive partitioning method.
Left unchecked, CART has the capability of placing each subject in
his own terminal node.  An important consideration therefore lies in
the selection of an appropriate number of splits (i.e., nodes). This
``pruning'' problem involves an inherent but familiar trade-off:
construct an accurate estimator that also avoids overfitting the model
to the data. CART uses cross-validation in combination with a
specified loss function in order to determine where to stop
partitioning the covariate space. In the resulting pruned tree, a
single predicted value is assigned to each terminal node. For example,
with a continuous outcome and an $L_2$ (i.e., squared-error) loss function, the
predicted value for each terminal node is the mean outcome for
all observations falling into that node.

\cite{Molinaro2010} recently developed \emph{partDSA}, a new
loss-based recursive partitioning method. Like CART, \emph{partDSA} divides
the covariate space into mutually exclusive and disjoint regions on
the basis of a chosen loss function.  The resulting regression models
continue to take the form of a decision tree; hence,
\emph{partDSA} also provides an excellent foundation for developing a
clinician-friendly tool for accurate risk prediction and
stratification. However, this algorithm
differs from CART in that the decision tree may be constructed
from both `and' and `or' conjunctions of predictors.  The advantage of
this representation is two-fold: (i) in cases where only 'and'
conjunctions of predictors are required to build a parsimonious model,
the \emph{partDSA} and CART representations coincide; and, (ii) in
cases where CART requires two or more terminal nodes to represent
distinct subpopulations (i.e., defined by covariates) having the same
outcome distribution, \emph{partDSA} can represent these same
structures using a single partition.  \cite{Molinaro2010} showed that this
increased flexibility  improves prediction accuracy and predictor
stability for uncensored continuous outcomes using a $L_2$
loss function in comparison to the best adaptive algorithms in the statistical
literature, including CART and Logic Regression \citep{LogicReg}.

In both CART and \emph{partDSA}, the choice of loss function plays a
key role in each step of the model building process. The default
choice for continuous outcomes (i.e., $L_2$ loss)
requires modification if these outcomes are censored.
Numerous adaptations of CART have been suggested in the
literature for handling right-censored outcomes; see, for example,
\cite{G&O_1985}, \cite{Ciampi_1986}, \cite{Segal_1988}
\cite{Therneau90}, \cite{L&C_1992}, and \cite{L&C_1993}.
With one exception, each of the aforementioned adaptations of CART
replaces the usual $L_2$ loss function with a criterion function that
relies on more traditional estimators and measures of fit used with
right censored outcome data.  In the absence of censoring, such
adaptations typically yield answers that differ from those provided by
the default implementation of CART.  Adaptations that replace the
$L_2$ loss function with other criteria also do not allow one to
easily quantify the prediction error of the final model using the same
loss function.  \cite{Molinaro_2004} proposed $CART_{IPCW}$, a direct
adaptation of CART that replaces the $L_2$ loss function with an
Inverse Probability Censoring Weighted estimator
\citep[IPCW;][]{R&R_1992} of the desired ``full data'' $L_2$ loss
function (i.e., that which would be used in the absence of censoring).
This allows for an otherwise unaltered implementation of CART to be
used for splitting, pruning, and estimation.  The IPCW-$L_2$ loss
function, computable for observed data, is under standard assumptions
a consistent estimator of the full data $L_2$ loss function.
\cite{Molinaro_2004} demonstrate that $CART_{IPCW}$ generally has
lower prediction error, measured as $L_1$ loss using the Kaplan Meier
median as the predicted value within a given node, when compared to
the one-step deviance method of \citet{L&C_1992}.

In this paper, we similarly extend
\emph{partDSA} to the setting of right-censored data using a
IPCW-$L_2$ loss function. The resulting methodology builds a decision tree
based on the (time-restricted) mean response in each node.
We also extend \emph{partDSA} to the problem of predicting a
binary outcome of the form $Z(t) = I( T \geq t ),$ where $T$ is the
event time of interest and $t$ is some specified time point.  This
extension involves a weighted modification of the $L_2$ loss
function that uses the {\em Brier Score} \citep{Graf1999}.
\cite{Graf1999} introduces
this measure as a way to compare the predictive accuracy
of various estimators of survival
experience.  Two novel contributions of this paper are to (i)
demonstrate that this score has a simple but interesting
representation as an IPCW-weighted $L_2$ function; and, (ii) make use
of this loss function as the basis for constructing a
\emph{partDSA} regression tree.

The remainder of this paper proceeds as follows. First, we describe
the relevant ``full data'' and ``observed data'' structures,
giving a brief overview of loss-based estimation in each
setting.  We then discuss how these ideas may be used to
extend the \emph{partDSA} algorithm to right-censored outcomes.
The performance of these extensions of \emph{partDSA} are evaluated
via simulations and in an analysis of  clinical trial data on brain
cancer patients.  Finally, we close the paper with a discussion and
comments on further, useful extensions.

\section{Methods}

\subsection{Relevant Data Structures}
\subsubsection{Full Data}

In the ideal setting, one observes $n$ i.i.d. observations $X_1,
\ldots , X_n$, of a ``full data'' structure, say $X=(T,W)$, where $T$
denotes a response variable and $W \in \mathcal{S}$ denotes a
(possibly high-dimensional) vector of covariates. Denote the
corresponding (unknown) distribution of $X$ by
$F_{X,0}$. With time-to-event data, our focus
hereafter, $T > 0$ is a continuous random variable that
denotes the event time of interest and $W$ represents a set of
baseline covariates. In this case, we may equivalently define the full
data as $X = (Z,W)$, where $Z = \log T$.  In the setting of
cancer risk prediction, $W$ may include age as well as various
genomic, epidemiologic and histologic measurements that are continuous
or categorical in nature.  More generally, the available
information may include both time-dependent and time-independent
covariates; we focus on the setting of time-independent $W$ only.

\subsubsection{Observed Data}
\label{sec: obs data}

In practice, information may be missing on one or more of the $X_is$; that
is, one instead observes $n$ i.i.d. observations $O_1, \ldots , O_n$
of an observed data structure $O$ having distribution $F_{Obs}$. With
time-to-event outcomes, the most common form of missing data
is right-censoring of event times due to
drop out or end of follow-up.  Here,
$O=\{\widetilde{T}=\min(T,C),\Delta = I(T \le C), W\},$ where
$\widetilde{T} = \min(T,C)$ and $\Delta = I(T \le C)$; equivalently,
$O = (\widetilde{Z},\Delta,X)$, where $\widetilde{Z} = \log
\widetilde{T}$.  The distribution $F_{Obs}$ of $O$ then depends on
$F_{X,0}$ and the conditional distribution
of the censoring variable $C$ given $X$, say $G_0( \cdot | X)$.

We assume that the censoring variable $C$ is conditionally
independent of $T$ given $W$; that is, $\bar{G}_0(\cdot|X) = \bar{G}_0(\cdot|W),$
where $\bar{G}_0(c \mid X) = Pr(C \geq c \mid X)$ and $\bar{G}_0(c \mid W)$ is defined
analogously. Since $X$ includes only
time-independent covariates, it therefore
satisfies a coarsening at random condition (CAR) \citep{Gilletal97}. In the absence
of CAR, fully-specified parametric models for the dependence
of $\bar{G}_0(\cdot|X)$ on $T$ (or $Z$) are needed and may be
used in combination with sensitivity analysis \citep[e.g.,][]{scharf02}.

\subsection{Loss-Based Estimation}
\subsubsection{Estimation with Full Data}
\label{sec: FDE}

Let $\psi: {\cal S} \rightarrow \mathbb{R}$ be a real-valued function
of $W$, where $\psi \in \Psi$. Let $L(X,\psi)$ denote a full
data loss function  and define $E_{F_{X,0}}\{L(X,\psi)\} = \int
L(x,\psi)dF_{X,0}(x)$ as the corresponding risk of $\psi$. The
parameter of interest, $\psi_0,$ is then defined as a minimizer of the
risk $E_{F_{X,0}}\{L(X,\psi_0)\} = \min \limits_{\psi \in
 \Psi}E_{F_{X,0}}\{L(X,\psi)\}.$
In practice, and on the basis of the fully observed data $X_1, \ldots
, X_n$, $\psi_0$ can be estimated by minimizing the empirical risk
(i.e., average loss) $n^{-1} \sum_{i=1}^n L(X_i, \psi)$.  Generally
speaking, feasible estimation procedures require $\psi(\cdot)$ to be
modeled in some fashion, in which case estimation of $\psi_0$ reduces
to estimating the corresponding model parameter(s). We reserve further
discussion of this issue until Section \ref{sec: pdsa}, where
piecewise constant regression estimators will be of primary
interest in connection with \emph{partDSA}.

The purpose of the loss function is to quantify a specific measure of
performance and, depending on the parameter of interest, there could
be numerous loss functions from which to choose. In the case of the
continuous outcome $Z = \log T$, the conditional mean $\psi_0(W) = E(Z \mid W)$
is frequently of interest. Under mild conditions,  $\psi_0(W)$
is the unique minimizer of the risk under
the squared error loss $L(X,\psi) = \{Z - \psi(W)\}^2$. However, it also
minimizes the risk under the much larger class of Bregman loss
functions \citep[e.g.,][]{bregman05}, with the corresponding optimal
risk measuring other aspects of performance.

In the case of time-to-event data, median survival and
estimated survivorship probabilities are often
of interest. It is easily shown that the conditional median
survival time $\psi_0(W)=Med(T \mid W)$ minimizes the risk under
the absolute error loss $L(X,\psi) = \mid T - \psi(W)\mid$.
Survivorship estimation can also be viewed as a loss
minimization problem. For example, let $Z(t) = I(T > t)$
denote survival status at a given time $t$. Then, the predictor $\psi_{t}(W)$
minimizing the risk under $L(X,\widetilde{\psi}_t)=\{Z(t)-\widetilde{\psi}_t(W)\}^2$
is $\psi_{0t}(W)=P(T>t|W)$ \citep[e.g.][]{Graf1999}.
Minimizing the corresponding average loss yields an estimator for $\psi_{0t}(W).$

\vspace*{-0.1in}
\subsubsection{Estimation with Observed Data}
\label{sec: ODE}

In the presence of right-censored outcome data, the goal remains to
find an estimator for a parameter $\psi_0$ that is defined in terms of
the risk for a full data loss function $L(X,\psi)$, e.g., a predictor
of the log survival time $Z$ based on covariates $W$. An immediate
problem is that $L(X,\psi)$ cannot be evaluated for any observation
$O$ having a censored survival time ($\Delta = 0$); for example,
$L(X,\psi) = \{Z - \psi(W)\}^2$ cannot be evaluated if $Z = \log T$ is
not observed. Risk estimators based on only uncensored observations,
such as $n^{-1} \sum_i L(X_i, \psi) \Delta_i$, are biased
estimators for $E_{F_{X,0}}\{L(X,\psi)\}$. Replacing the full (uncensored)
data loss function by an observed
(censored) data loss function having the same risk
provides one possible solution to this problem.

IPCW estimators with right-censored outcomes
were introduced in \cite{R&R_1992}. Though most frequently used
in the development of unbiased observed data estimating equations, the
IPCW principle is valid in great generality and can be used to
solve the desired risk estimation problem.
Assume $E_{F_{Obs}}(\Delta|X) = \bar{G}_0(T|W) > 0,$
where $\bar{G}_0(c \mid W) = Pr(C \geq c \mid W)$; see Section
\ref{sec: obs data}. Then,  $E_{F_{Obs}}\{\Delta L(X,\psi)
/ \bar{G}_0(T \mid W) \} = E_{F_{X,0}}\{ L(X,\psi)\},$ implying that
$n^{-1} \sum_{i=1}^n \Delta_i L(X_i,\psi) / \bar{G}_0(T_i \mid W_i)$
is an unbiased estimator of the desired full data risk $E_{F_{X,0}}\{
L(X,\psi)\}$.  A simple but important example is the IPCW version of
the $L_2$ loss function,
given by the previous formula with $L(X_i,\psi) = \{Z_i - \psi(W_i)\}^2$.

In general, the censoring probability function $\bar{G}_0(\cdot|W)$ is
unknown and must be estimated from the data. For any
consistent estimator $\bar{G}_n(\cdot|W)$ of $\bar{G}_0(\cdot|W),$
\begin{eqnarray}
\label{e: ipcw loss}
\frac{1}{n} \sum^n_{i=1} L(X_i, \psi) \frac{
 \Delta_i}{\bar{G}_n(T_i|W_i)}
\end{eqnarray}
is under mild conditions a consistent estimator for the
full data risk $E_{F_{X,0}}\{L(X,\psi)\}.$ Moreover, in the absence of
censoring, \eqref{e: ipcw loss} reduces to the desired
empirical risk. As such, it is reasonable to estimate $\psi(W)$
by minimizing \eqref{e: ipcw loss}.  One can also estimate
$\bar{G}_0(\cdot|\cdot)$ using covariates other than those used for
estimating $\psi(W)$, allowing for
certain forms of informative censoring.  An
important drawback
of \eqref{e: ipcw loss} is the need for
$\bar{G}_n(\cdot|W)$ to be consistently estimated.

\subsection{The Brier Score}

As suggested in Section \ref{sec: FDE}, the prediction of survival
status at a given time $t$ may be viewed as a loss minimization
problem: estimate $\psi_t(W)$ by minimizing the average full
data loss function $n^{-1} \sum_{i=1}^n L(X_i,\psi),$ where
$L(X_i,\psi) = \{Z_i(t) - \psi(W_i)\}^2$ and $Z_i(t) = I( T_i > t )$. In
\cite{Graf1999}, this empirical loss function is referred to as the
Brier score, and is proposed as a measure of prediction
inaccuracy that permits comparisons between competing models for
$\psi_{0t}(W)=P(T>t|W)$.
\cite{Graf1999} also extend the definition of the Brier score to
accommodate censored survival data. In particular, given a
time $t$, they propose to divide the observations into three groups.
Group 1 contains subjects censored before $t$; here, $\Delta_i=0,$
$\widetilde{T}_i \leq t$ and $Z_i(t)$ cannot be determined.  Group 2 contains
subjects experiencing an event before $t$;
here, $\Delta_i=1$, $\widetilde{T}_i \leq t$, and $Z_i(t) = 0$.
Group 3 contains subjects that remain at risk for
an event at time $t$; here, the value of $\Delta_i$
is irrelevant, for $\widetilde{T}_i>t$ and thus $Z_i(t) = 1$.
The corresponding average observed data loss function,
generalized here to allow for covariate-dependent censoring
and assuming $t \neq \tilde{T}_i$ for any $i$,
is computed as follows:
\begin{equation}
\label{orig B score}
\begin{split}
BS^c(t)=\frac{1}{n} \sum_{i=1}^{n} \left [ \{0-\hat{\psi}_t(W_i)\}^2
\times \frac{I(\widetilde{T_i} \le
t,\Delta_i=1)}{\bar{G}_n(\widetilde{T}_i|W_i)}
+\{1-\hat{\psi}_t(W_i)\}^2 \times \frac{I(\widetilde{T}_i>t)}{\bar{G}_n(t|W_i)}
\right ]
\end{split}
\end{equation}
\noindent
where $I(\widetilde{T_i} \le t,\Delta_i=1)$ is the indicator function for
Group 2, $I(T_i>t)$ is the indicator function for Group 3, and
$\bar{G}_n(\cdot|W)$ estimates the censoring
survival function $\bar{G}_0(\cdot|W) = P(C \geq \cdot|W)$.  Observe that
subjects in Group 1 do not contribute to this loss function
except through estimation of $\bar{G}_0(\cdot|W).$

We now demonstrate that $BS^c(t)$ has an interesting IPCW
representation.
Proceeding similarly to \cite{straw2000}, define
$\tilde{T}_i(t) = \min(T_i(t),C_i)$ and
$\Delta_i(t) = I\{ T_i(t) \leq C_i\},$ where
$T_i(t) = \min(T_i,t)$. Clearly,
$\Delta_i(t) \rightarrow \Delta_i$ as $t \rightarrow \infty$
and it is evident that any subject with $\Delta_i = 1$ must also
have $\Delta_i(t) = 1$ for every $t$. Thus, the importance of
$\Delta_i(t)$ only becomes apparent when $\Delta_i = 0$; in
particular, for a given $t < \infty$, it is possible for $\Delta_i =
0$ and $\Delta_i(t) = 1$.  Specifically, for a subject in Group 1,
since $\Delta_i=0$ and $\widetilde{T}_i \leq t$, we have $C_i \leq t$
and $C_i < T_i$. It follows that $\tilde{T}_i(t) = C_i$ and
$\Delta_i(t) = 0$ (i.e., except if $t = C_i$, which is
impossible if $t \neq \tilde{T}_i$ for any $i$).
For a subject in Group 2, since $\Delta_i=1$ and
$\widetilde{T}_i \leq t$, we have $T_i \leq C_i$ and $T_i \leq t$.  It
follows that $I(\widetilde{T_i} \le t,\Delta_i=1) = \Delta_i(t) = 1$
and that $\widetilde{T}_i(t) = T_i$. For a subject
in Group 3, we have $\widetilde{T}_i>t$ regardless of $\Delta_i$ and
therefore that $T_i > t$ and $C_i > t$.  It follows that
$I(\widetilde{T_i} > t) = \Delta_i(t) = 1$ and that
$\widetilde{T}_i(t) = t$. It is then easy to show
that \eqref{orig B score} may be written as:
\begin{eqnarray}
\label{e: modBS}
BS^c(t) & = &  \frac{1}{n} \sum_{i=1}^{n}
\frac{\Delta_i(t)}{\bar{G}_n\{T_i(t) | W_i\}}
\times \left\{ Z_i(t)-\hat{\psi}_t(W_i) \right\}^2.
\end{eqnarray}
The IPCW-like risk estimator $BS^c(t)$ uses a modified event time and
censoring indicator and will be consistent for the expected (full
data) Brier score under mild regularity conditions.  Such a loss
function is also easily extended to the setting of multiple time
points; for example, estimators for $\psi_0(W)$ might be obtained by
minimizing loss functions of the form $\sum_{r=1}^p BS^c(t_r), $
$\int_0^{\tau} BS^c(t) w(t) dt,$ or even $\max \{ BS^c(t_1),
BS^c(t_2), \ldots, BS^c(t_p) \}.$
In comparison with estimators derived from the $L_2$ loss function,
loss functions derived from the Brier score have a nice invariance
property, being unaffected by monotone transformations
of the time-to-event variable (i.e., whether or not censoring is present).


\subsection{partDSA: recursive partitioning for
full data structures and extensions}

\label{sec: pdsa}

\emph{partDSA} \citep{Molinaro2010} is a statistical methodology for
predicting outcomes from a complex covariate space with fully observed
data. Similarly to CART, this novel algorithm generates a piecewise
constant estimation list of increasingly complex candidate predictors
based on an intensive and comprehensive search over the entire
covariate space. A brief description is below;
see \citet{Molinaro2010} for further details.

The strategy for estimator construction, selection, and performance
assessment in \emph{partDSA} is entirely driven by the specification
of a loss function and involves three main steps. Step 1 involves
defining the parameter of interest as the minimizer of an expected
loss function (i.e. risk), where the given loss function represents a
desired measure of performance. In Step 2, candidate estimators are
constructed by minimizing the corresponding empirical risk (i.e.
average loss) function.  For this reason, the regression function
$\psi(W)$ should ideally be parameterized in a way that generates a
simple estimator.  \emph{partDSA} relies on piecewise constant regression
models to approximate the desired parameter space
and uses three types of operations to build these models
(illustrated in Web Appendix A).  Finally, in Step
3, cross-validation is used to estimate risk and to select an optimal
estimator among the candidate estimators obtained in Step 2
\citep[e.g.,][]{V&D_2003}.  The \emph{partDSA} software, currently
implemented for problems without missing data for select loss
functions (e.g., $L_2$) is available on CRAN as an \texttt{R} package
\citep{Molinaro_2009}.

The results summarized in Section \ref{sec: ODE} demonstrate how the
problem of estimating a parameter that minimizes the full data risk
under a given loss function can be generalized to right-censored outcome data:
replace the desired average full data
loss with the corresponding average observed data IPCW-weighted
loss. Hence, the $L_2$-loss-based estimation capabilities of
\emph{partDSA} may be extended to right-censored outcomes
in the following two ways: (i) $\emph{partDSA}_{IPCW},$ that is,
\emph{partDSA} implemented using the average loss function \eqref{e:
 ipcw loss} with $L(X_i,\psi) = \{Z_i - \psi(W_i)\}^2.$
and, (ii) $\emph{partDSA}_{Brier}$, that is, \emph{partDSA}
implemented using the weighted $L_2$ loss function \eqref{e: modBS}
for the binary outcome $Z_i(t)$.
Notably, the first extension generalizes the Koul-Susarla-van Ryzin estimator
for the accelerated failure time model \citep{KSV}
allowing for covariate-dependent censoring and a tree-based regression function.



\section{Simulation Studies} \label{s:sims}
We ran four simulation studies to evaluate the performance
of $\emph{partDSA}_{IPCW}$ and several variations on the Brier score
loss function. In addition to these adaptations of \emph{partDSA},
we include for the sake of comparison the methods of \citet[$L\&C_{NLL}$]{L&C_1993}
and the IPCW-based extension of CART developed in \citet[$CART_{IPCW}$]{Molinaro_2004}.
The variations on the Brier score loss function include $BS^c(t)$
($1 Fixed$, with $t$ set equal to a fixed time point)
and two aggregate loss functions, one based on five evenly spaced time points
($5 Even$) and the other using five percentiles derived from
the Kaplan-Meier estimate of the marginal survival distribution ($5 KM$).
Further details on these loss function specifications and algorithm
tuning parameters may be found in Web Appendix B.

The structure of each simulation study is the same. First, a training set
of 250 observations was generated with a nominal censoring level ($0\%$, $30\%$ or $50\%$);
a model size (i.e., number of terminal partitions) is then determined using the first
minimum of the 5-fold cross-validated risk.  The performance of this
model is then assessed with
an independent test set of 5000 observations generated from the full data distribution
described in Section \ref{TSEC} (i.e. no censoring). This process was repeated for 1000
independent (training, test) set combinations.

Each simulation study involves five covariates $W_1 \ldots W_5$, each
of which is a discrete uniform variable on the integers 1-100. Only $W_1$ and $W_2$
influence survival times; these are generated from an exponential distribution with a covariate-dependent mean parameter $\sigma$. We consider both a ``high'' and `low'' signal setting.
In the high signal scenario, $\sigma$ is set equal to 5 if $W_1 > 50$ or $W_2 > 75$ and
0.5 otherwise; in the low signal scenario, the values for $\sigma$ are 1 and 0.5 respectively.

\begin{figure}[!htb]
\begin{center}
\includegraphics[width=6in]{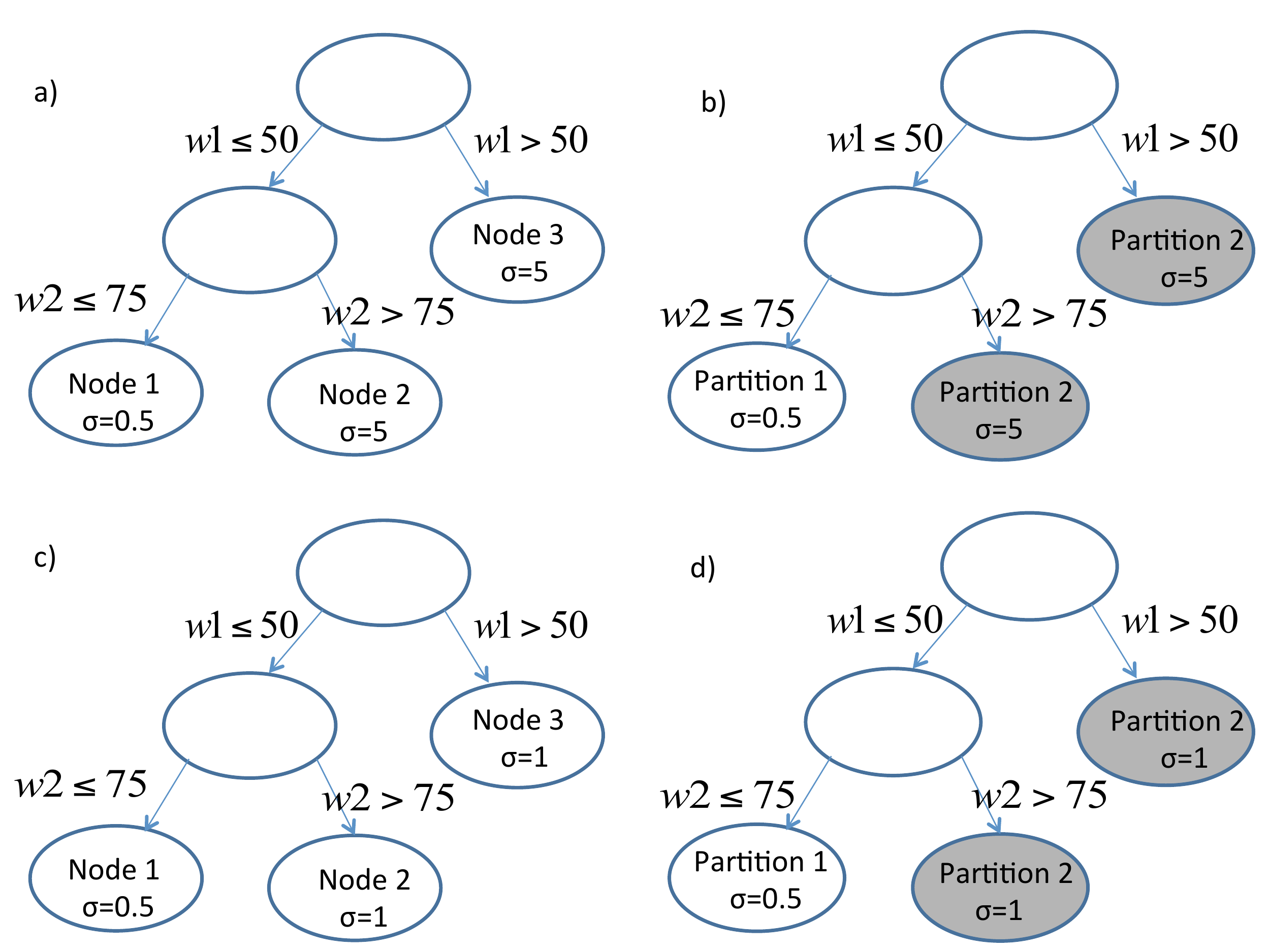}
\caption{{\em True Models for Multivariate Simulations of Section
\ref{s:MVsims}.}
Panel (a): High Signal: CART representation.
Panel (b): High Signal: \emph{partDSA} representation.
Panel (c): Low Signal: CART representation.
Panel (d): Low Signal: \emph{partDSA} representation.}
\label{Fig:MultiTT}
\end{center}
\end{figure}

Importantly, there is a single true regression model in each study having
two correct regression tree representations. The true CART
tree structure has three terminal nodes; however,
two of these nodes represent distinct subpopulations of
subjects  {\em having the same survival distribution}. In contrast, the true
\emph{partDSA} tree has two terminal partitions,
combining these two terminal nodes into a single partition (i.e.,
a more parsimonious representation). Figure \ref{Fig:MultiTT} summarizes
these representations for both the high and low signal settings.

In each study, censoring times are independently generated from uniform
distributions. For each signal level, two censoring settings are considered.
In the first setting, censoring is allowed to depend on covariate information, such
that an (approximately) equal level of censoring exists in both risk groups that
is close to the specified nominal level. In the second setting, censoring is
generated independently of all covariate information, such that overall censoring levels
approximately match the specified nominal level. Censoring probabilities used in the
calculation of censoring weights are respectively estimated using survival curves derived
from an incorrectly specified Cox regression model (i.e., the correct
variables are included, but not modeled correctly) and a properly specified
nonparametric product-limit estimate of survival.
To ensure these estimated censoring probabilities
remain bounded away from zero, observed survival times are truncated using
a sample-dependent truncation time, as described in Web Appendix B. The actual
censoring level decreases slightly after truncation; therefore, both nominal and
empirical levels are reported.

Below, we only report results for the setting
of covariate-dependent censoring in Section \ref{s:MVsims}, where censoring
probabilities are estimated using Cox regression. Important performance differences
in the setting of covariate-independent censoring are highlighted;
figures and tables referenced  but not found in the main document may be found
in Web Appendix C. Notably, since $CART_{IPCW}$ and $\emph{partDSA}_{IPCW}$ use the
same loss function, a direct comparison of these results illustrates the
difference between the \emph{partDSA} and \emph{CART} algorithms,
controlling for both censoring type and level. However,
before giving these results, we first describe the metrics that will be
used to evaluate prediction performance in the test set.

\subsection{Test Set Evaluation Criteria}
\label{TSEC}

Performance evaluations are based on four criteria:
prediction concordance, prediction error, proper risk
stratification, and pairwise predictive similarity. In each case,
we intend to evaluate how well a model built using censored data performs
on an independent, fully observed  test set (i.e., no censoring).
For brevity, we use ``tree'' and ``terminal node'' to describe the structure
and output for both \emph{partDSA} and CART in this section, with the
understanding that the true meanings do differ
somewhat for \emph{partDSA}.
Below, a brief description of each measure is given;
a detailed description of each measure may be found in
Web Appendix C.

\emph{Prediction Concordance ($C_p,~ \bar{C}_p$):}
To ensure comparability across the different methods of tree
construction, we define prediction concordance using
the terminal-node-specific IPCW-estimated average survival time derived
from the training set data as the predicted outcome. Observed
and predicted outcomes for subjects in the test set are then
compared using the concordance index $C_p$ \citep{Harrell_1982}
and a modification $\bar{C}_p$ suggested in \cite{YanGreene08}
that typically exhibits less bias in settings
where ties between predicted values are possible. Values
near 1.0 (0.5) indicate excellent (poor) concordance.

\emph{Prediction Error ($L_p$):} In this method of evaluation,
we compare for each method the predictions for each test
set subject obtained using the true tree structure
(cf.\ Figure \ref{Fig:MultiTT}) and the corresponding
estimated tree structure. The average squared error
of the difference in these predictions, $L_p,$
equals zero if and only if both trees classify all test set
subjects into the same risk groups, and increases away from zero as
the heterogeneity in risk group assignment, hence predicted risk, increases.

\emph{Pairwise Prediction Similarity ($D_p$):} This criterion,
motivated by work in \cite{Chip01},
targets the ability of each method to identify groups of subjects having a
similar level of risk. In particular, considering all possible
pairings of subjects, this measure looks at the extent to which
each true tree and corresponding set of estimated trees
classify pairs of subjects into the same risk groups.
With perfect agreement, $D_p = 1$; with perfect disagreement, $D_p = 0$.
An advantage of this metric is its independence from both tree topology and
the actual predictions associated with each terminal node.

\emph{Risk Stratification:} This criterion also focuses on
the ability to properly separate patients into groups of
differing risk.  In particular, for each of the 1000 independent test
sets, node-specific empirical survivor functions are computed.  Then,
for each estimation method, and for the subset of the 1000 estimated
trees that consist of either two or three terminal nodes, we compute
the corresponding empirical survivorship and 0.025 and
0.975 percentiles on a fine grid of time points. A graphical summary
of the results is provided for each simulation study.
As suggested by Figure \ref{Fig:MultiTT},
we expect to see a high proportion of
cases for \emph{partDSA} (CART) with only two (three) survival
curves and smaller (larger) standard errors.

\subsection{Results}

\label{s:MVsims}
\label{s:MVsims1}

As seen  in Table \ref{Table:MultiSim1_1}, all methods build slightly
larger models than necessary while selecting the two signal variables consistently.
Overall, $\emph{partDSA}_{Brier}(5 even)$ and $\emph{partDSA}_{IPCW}$ exhibit the
best performance on the evaluation metrics introduced in Section \ref{TSEC},
with $\emph{partDSA}_{IPCW}$ also consistently outperforming $\emph{CART}_{IPCW}.$

In general, c-indices are comparable, with the \emph{partDSA} methods doing as well,
and usually better, than the CART-based methods. The relative improvement
in performance based on the mean c-index ranges up to 6\%, with the greatest
improvement being seen at the highest censoring level.
In addition, $\emph{partDSA}_{IPCW}$
and $\emph{partDSA}_{Brier}(5KM)$ have the lowest prediction error in the presence
of censoring, with $\emph{partDSA}_{IPCW}$ having the lowest prediction error for $0\%$
and $30\%$ censoring and $\emph{partDSA}_{Brier}(5KM)$ having the lowest error
in the case of $50\%$ censoring.  In comparison with the
CART-based methods, the relative decrease in prediction error ranges up to $53\%.$
In comparison with CART-based methods, the \emph{partDSA} methods also perform
approximately $15\%$ better on the pairwise prediction similarity measure.
Web Figures ~\ref{WebFig:KMMulti10}-\ref{WebFig:KMMulti150}
depict the average empirical survival curves for the four methods,
where the \emph{partDSA} and CART methods routinely identify two
and three distinct groups of patients, respectively; see
Web Table \ref{WebTable:MV1_1ModelSize}).
However, in the case of CART, the clinical utility of these
three risk groups is not as apparent, for two of these groups are
estimated to have nearly identical survival experiences, each having
wider standard errors.

In general, some variation in performance is observed across the different choices of
Brier score loss function, with $\emph{partDSA}_{Brier}(5KM)$ consistently
performing well on all four metrics, followed by $\emph{partDSA}_{Brier}(1 Fixed)$ and then
$\emph{partDSA}_{Brier}(5even).$ The difference in performance between
$\emph{partDSA}_{Brier}(5KM)$ and $\emph{partDSA}_{Brier}(5even)$ is noticeable and
cannot be explained by the introduction of censoring, for the performance of both is
reasonably stable for each metric across censoring levels. This is less true for
$\emph{partDSA}_{Brier}(1 Fixed),$ where more substantial changes in
prediction error and pairwise predictive similarity are observed
with increasing levels of censoring.

\begin{table}
  \caption{Simulation Results: High Signal, Covariate-Dependent Censoring. Reported are true and average fitted model sizes, number of predictors
    in the fitted model,  number of correct (incorrect) predictors selected, c-index ratios ($C_p$, $\bar{C}_p$; larger values indicate better performance), prediction error ($L_p$; smaller values indicate less error), and pairwise prediction similarity error ($D_p$; larger values indicate less error) for the four methods over three censoring levels for the simulation.
    The $1 Fixed$ method refers to $\emph{partDSA}_{Brier}$ using a single fixed time point;  the $5 even$ method refers to $\emph{partDSA}_{Brier}$ using 5 evenly spaced time points; and, the $5 KM$ method refers to $\emph{partDSA}_{Brier}$ using 5 time points determined using percentiles of the Kaplan-Meier estimate of the marginal survivor function.}

\begin{center}
\begin{tabular}{ c  c | c  c  c | c  c | c} 

\multicolumn{2}{c}{    }  &   \multicolumn{3}{c}{$\emph{partDSA}_{Brier}$}  &     \multicolumn{2}{c}{IPCW}     & \\ \hline
Censoring &Criteria & $1 Fixed$ & $5 Even$ & $5 KM$ & $\emph{partDSA}$ & $CART$ & $L\&C_{NLL}$ \\  \hline 
\hline 
& True Model Size & 2.00 & 2.00 & 2.00 & 2.00 & 3.00 & 3.00 \\ \hline
\multirow{4}{*}{$0\%$/$0\%$}
&Fitted Size&2.198&2.450&2.228&2.170&3.106&3.192\\
&\# Predictors&2.135&2.095&2.195&2.111&2.047&2.076\\
&\# W1-W2&1.998&1.809&2.000&1.999&1.962&1.996\\
&\# W3-W5&0.137&0.286&0.195&0.112&0.085&0.080\\
&$C_p$ &0.879&0.828&0.880&0.891&0.812&0.809\\
&$\bar{C}_p$&0.782&0.748&0.783&0.789&0.748&0.747\\
&$L_p$ &0.082&0.270&0.076&0.061&0.076&0.072\\
&$D_p$ &0.944&0.838&0.946&0.964&0.845&0.842\\
\hline
\multirow{4}{*}{$30\%$/$26.4\%$}
&True Size&2.000&2.000&2.000&2.000&3.000&3.000\\
&Fitted Size&2.249&2.391&2.227&2.230&3.058&3.222\\
&\# Predictors&2.190&2.160&2.206&2.146&1.988&2.090\\
&\# W1-W2&1.993&1.894&2.000&1.999&1.904&1.988\\
&\# W3-W5&0.197&0.266&0.206&0.147&0.084&0.102\\
&$C_p$ &0.867&0.839&0.878&0.881&0.810&0.802\\
&$\bar{C}_p$&0.775&0.756&0.783&0.784&0.744&0.742\\
&$L_p$ &0.125&0.225&0.084&0.082&0.126&0.118\\
&$D_p$ &0.918&0.857&0.940&0.944&0.828&0.825\\
\hline
\multirow{4}{*}{$50\%$/$45.7\%$}
&True Size&2.000&2.000&2.000&2.000&3.000&3.000\\
&Fitted Size&2.396&2.361&2.251&2.359&2.937&3.093\\
&\# Predictors&2.137&2.136&2.200&2.191&1.875&1.983\\
&\# W1-W2&1.908&1.888&1.997&1.974&1.802&1.881\\
&\# W3-W5&0.229&0.248&0.203&0.217&0.073&0.102\\
&$C_p$ &0.845&0.844&0.874&0.864&0.810&0.794\\
&$\bar{C}_p$&0.761&0.759&0.781&0.775&0.740&0.732\\
&$L_p$ &0.199&0.216&0.102&0.125&0.188&0.215\\
&$D_p$ &0.862&0.855&0.926&0.906&0.805&0.792\\
\hline
\end{tabular}
\end{center}
\label{Table:MultiSim1_1}
\end{table}

There is greater heterogeneity in the performance comparison for the case of
covariate-independent censoring; see Web Table ~\ref{WebTable:MultiSim1_1nonInf}.
Noteworthy observations include (i) a consistent dominance of the CART-based method $L\&C_{NLL}$
on the prediction error metric in the presence of censoring; (ii) the
consistent and generally high level of performance of $\emph{partDSA}_{Brier}(5KM)$
on all metrics; and, (iii) the continued dominance of $\emph{partDSA}_{IPCW}$ over
$\emph{CART}_{IPCW}$ on all metrics regardless of censoring level, despite a more
noticeable degradation in performance on the prediction error and pairwise predictive
similarity indices as censoring increases in comparison with Table \ref{Table:MultiSim1_1}.


The results of the low signal simulation are summarized in Table \ref{Table:MultiSim1_2}.
All methods select models less than the ideal size, choosing fewer correct W1-W2 variables,
though still with greater frequency in comparison with the incorrect W3-W5 variables. Further information on chosen model sizes may be found in Web Table \ref{WebTable:MV1_2ModelSize}.
In general, the partDSA methods perform similarly to the CART methods on all four performance
metrics; however, relative to CART, moderate improvement is observed at the 50\% censoring level
on all metrics for $\emph{partDSA}_{IPCW}$ and $\emph{partDSA}_{Brier}(5Even).$ Empirical
survival curves are summarized in Web Figures \ref{WebFig:KMMulti1Low0}--\ref{WebFig:KMMulti1Low50}, where patterns are generally similar to those reported earlier. Similar results are observed
in the case of covariate-independent censoring, though the improvement observed at the 50\%
censoring level in the high signal setting is no longer evident; see Web Tables \ref{WebTable:MultiSim1_2nonInf} and \ref{WebTable:MV1_2nonInfModelSize} and
Web Figures \ref{WebFig:KMMulti1LownonInf0} --\ref{WebFig:KMMulti1LownonInf50}.

\begin{table}
\caption{Simulation Results: Low Signal, Covariate-Dependent Censoring.
 Reported are true and average fitted model sizes, number of predictors
    in the fitted model,  number of correct  predictors selected, c-index ratios ($C_p$, $\bar{C}_p$; larger values indicate better performance), prediction error ($L_p$; smaller values indicate less error), and pairwise prediction similarity error ($D_p$;
    larger values indicate less error) for the four methods over three censoring levels for the simulation.
    The $1 Fixed$ method refers to $\emph{partDSA}_{Brier}$ using a single fixed time point;  the $5 even$ method refers to $\emph{partDSA}_{Brier}$ using 5 evenly spaced time points; and, the $5 KM$ method refers to $\emph{partDSA}_{Brier}$
    using 5 time points determined using percentiles of the Kaplan-Meier
    estimate of the marginal survivor function.}

\begin{center}

\begin{tabular}{ c  c | c  c  c | c  c | c} 

\multicolumn{2}{c}{    }  &   \multicolumn{3}{c}{$\emph{partDSA}_{Brier}$}  &     \multicolumn{2}{c}{IPCW}     & \\ \hline
Censoring &Criteria & $1 Fixed$ & $5 Even$ & $5 KM$ & $\emph{partDSA}$ & $CART$ & $L\&C_{NLL}$ \\  \hline 
\hline 

 & True Model Size & 2.00 & 2.00 & 2.00 & 2.00 & 3.00 & 3.00 \\ \hline
\multirow{4}{*}{$0\%$/$0\%$}
&Fitted Size&1.642&1.502&1.389&1.456&1.590&1.873\\
&\# Predictors&0.705&0.521&0.496&0.586&0.571&0.838\\
&\# W1-W2&0.621&0.471&0.410&0.491&0.527&0.786\\
&\# W3-W5&0.084&0.050&0.086&0.095&0.044&0.052\\
&$C_p$ &0.608&0.604&0.602&0.607&0.603&0.610\\
&$\bar{C}_p$&0.567&0.562&0.559&0.563&0.563&0.571\\
&$L_p$ &0.080&0.086&0.091&0.087&0.083&0.071\\
&$D_p$ &0.618&0.592&0.581&0.600&0.608&0.644\\
\hline
\multirow{4}{*}{$30\%$/$26.7\%$}
&Fitted Size&1.550&1.536&1.290&1.456&1.571&1.546\\
&\# Predictors&0.676&0.648&0.427&0.705&0.553&0.523\\
&\# W1-W2&0.575&0.527&0.319&0.551&0.517&0.473\\
&\# W3-W5&0.101&0.121&0.108&0.154&0.036&0.050\\
&$C_p$ &0.607&0.604&0.597&0.611&0.607&0.607\\
&$\bar{C}_p$&0.565&0.562&0.554&0.565&0.565&0.563\\
&$L_p$ &0.073&0.076&0.084&0.076&0.073&0.075\\
&$D_p$ &0.607&0.591&0.565&0.605&0.609&0.597\\
\hline
\multirow{4}{*}{$50\%$/$45.7\%$}
&Fitted Size&1.648&1.780&1.306&1.591&1.572&1.282\\
&\# Predictors&0.812&1.106&0.533&0.943&0.552&0.264\\
&\# W1-W2&0.659&0.911&0.375&0.758&0.513&0.233\\
&\# W3-W5&0.153&0.195&0.158&0.185&0.039&0.031\\
&$C_p$ &0.607&0.616&0.602&0.614&0.607&0.601\\
&$\bar{C}_p$&0.566&0.574&0.556&0.570&0.565&0.555\\
&$L_p$ &0.056&0.049&0.066&0.055&0.058&0.067\\
&$D_p$ &0.609&0.649&0.571&0.629&0.608&0.562\\
\hline
\end{tabular}
\end{center}
\label{Table:MultiSim1_2}
\end{table}

\section{Data Analysis}
 \label{s:DA}

We now illustrate the model building capabilities of \emph{partDSA}- and
CART-based methods using data from 12 North American Brain Tumor Consortium (NABTC)
Phase II clinical trials for recurrent glioma \citep{Wu01022010}, run to assess the
efficacy  of novel therapeutic agents in patients with grade III or IV gliomas.
As there are few efficacious treatments for high-grade glioma and median survival is low
\citep[14.6 months;][]{doi:10.1056/NEJMoa043330}, there is strong interest in
identifying prognostic factors associated with overall survival.
Recently, several studies have combined data from multiple (necessarily small)
Phase II clinical trials to examine such factors using a various statistical methods; however,
findings have at best been moderately consistent \citep{Wong01081999, Carson20062007, Wu01022010}.

\citet{Wu01022010} combined the NABTC data with 15 North Central Cancer Treatment Group (NCCTG) trials and found tumor grade, patient age, baseline performance score, and time since diagnosis as important prognostic variables. Based on these results and those of earlier studies,
\citet{Wu01022010} conclude that there is strong evidence to suggest
that future trials should collect and report this information
as predictors of patient prognosis. Limitations of these analyses, and those to be presented below, include the possibility of trial referral bias and biases induced as a result of variation in patient eligibility criteria across studies.

The NABTC data set analyzed here includes 549 patients that were treated on Phase II trials between February 1998 and December 2002 \citep{Wu01022010}; Web Table \ref{T:Glioma} summarizes the data on 18 variables that include age, gender and several variables documenting a patient's health and tumor status as well as current/past therapies.
The outcome of interest is overall survival (OS), defined as time from study registration date to the date of death due to any cause (median OS was 30.4 weeks). Thirty patients were censored at last follow-up date, being still alive or lost to follow-up.
There are several noteworthy differences between the analysis and results here and in \cite{Wu01022010}. First, many of the results in that paper refer to the combined analysis of 27 trials, not just the 12 trials considered here; in addition, and important to the motivation for our analysis, is the fact that the risk groups ultimately identified in \citet[Table 5]{Wu01022010} were determined
by post-processing the results of their analyses in a subjective and relatively ad hoc manner.
In particular, their primary analysis consisted of estimating the cumulative hazard function
for OS using a Cox proportional hazard model adjusted only for current temozolomide (TMZ) use
and then subsequently analyzing this predicted count using CART without any further accounting
for censoring. In subsequent analysis, logrank tests were used to test for pairwise differences
between terminal nodes; terminal nodes were then combined for the purposes of defining risk groups if the p-value  from a corresponding log-rank test was $>0.01$ \citep[][Table 5]{Wu01022010}.  That is, \cite{Wu01022010} form risk groups
using a combination of 'and' statements derived from CART and 'or' statements
created using the indicated testing procedure. Important differences between
their approach and \emph{partDSA} include an improper accounting for censoring,
a lack of objectivity in the methodology used for defining risk groups,
and the failure to use cross-validation in evaluating the final models.

\begin{figure}[!htb]
\label{Fig: KM NABTC}
\begin{center}
\includegraphics[width=6in]{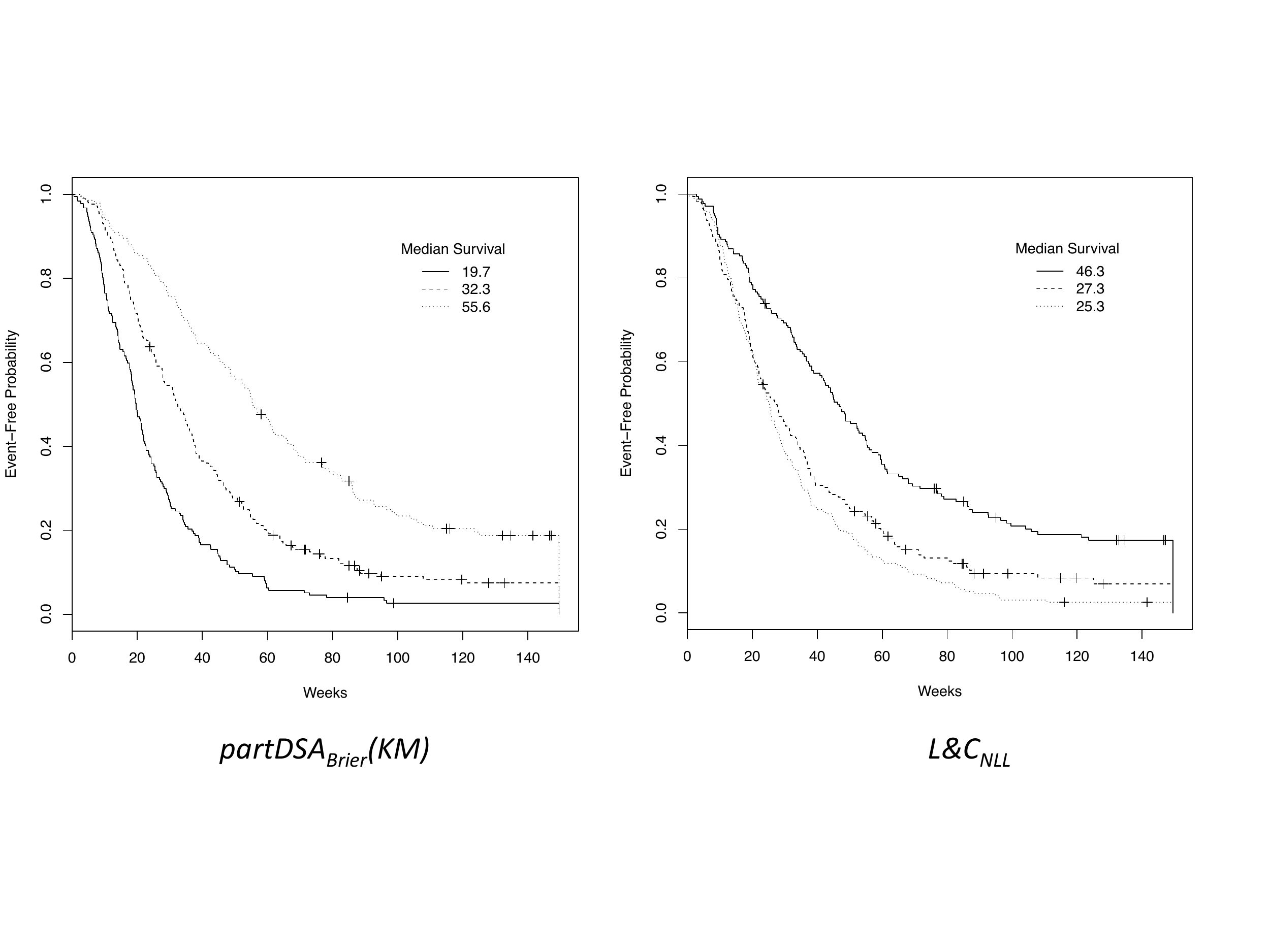}
\caption{{\em Kaplan-Meier Curves for NABTC data analysis in Section
\ref{s:DA}}.} Left panel: $partDSA_{Brier}(5KM)$  stratification of patients into three risk groups.
Right panel: $L\&C_{NLL}$   stratification of patients into three risk groups.

\label{Fig:KMCurves}
\end{center}
\end{figure}

All four algorithms from Sect.~\ref{s:sims} were run using 10-fold
cross-validation; below, we summarize the results from $partDSA_{Brier}(5KM)$
and $L\&C_{NLL}$, with further discussion of the remaining results
in Web Appendix D.
The risk stratification determined by $partDSA_{Brier}(5KM)$
is based on Karnofsky performance score (KPS; a marker for overall tumor burden and
cumulative treatment-related toxicity), patient age,  time since diagnosis,
baseline steroid use, gender, prior TMZ therapy (an approved
treatment for recurrent gliomas), and current TMZ therapy.
Three primary risk groups are identified; these are depicted in Table
\ref{Table: partDSA Brain}, with corresponding survival experiences
summarized by Kaplan-Meier curves in the left panel
of Figure \ref{Fig: KM NABTC}.
The low, medium and high risk groups
respectively have median survival times of 55.6 weeks (144 patients),
32.3 weeks (218 patients), and 19.7 weeks (187 patients); see Table
\ref{Table: partDSA Brain}.
In a sub-analysis of the NABTC data, \citet[Table 7]{Wu01022010}
found the same first five variables using a predicted outcome
variable that adjusts for current TMZ therapy. Hence, the only difference
in the variables selected is the inclusion of prior TMZ use; this clinically
relevant variable indicates that a patient  previously failed TMZ therapy,
lessening the chances of successful treatment.

For comparison, Table \ref{Table: LC-CART Brain} and Figure \ref{Fig: KM NABTC}
(right panel) summarize the corresponding results for $L\&C_{NLL}$.
It is observed that $L\&C_{NLL}$ selects only two of the variables identified by
$partDSA_{Brier}(5KM)$ (i.e. age split at 55 and prior TMZ),
but creates three risk groups. The lowest risk group
is defined by younger patients without prior TMZ treatment (median survival of 46.3 weeks,
176 patients); the highest risk group is defined by older patients (median
survival of 25.3 weeks, 195 patients). The two groups at higher risk
are also observed to have very similar survival experiences,
and the estimated spread in median survival times is approximately 15 weeks
shorter for $L\&C_{NLL}.$ With the exception of the low risk group, the
reported 95\% confidence intervals for median survival times
associated with the Kaplan-Meier curves for $partDSA_{Brier}(5KM)$ are
also considerably tighter than those associated with the risk groups identified
by $L\&C_{NLL}.$ Arguably, the separation
in risk identified here is not as clinically relevant as that found by \emph{partDSA}.
We note here that the 95\% confidence intervals presented in these two
tables are only meant to provide a sense of variability; because of the
post-hoc manner in which they are obtained, such intervals are not expected
to have the correct frequentist coverage.
Further comments and comparisons to results in
\cite{Wu01022010} may be found in Web Appendix D.

\begin{table}
 \caption{$partDSA_{Brier}(5KM)$ stratification   of NABTC patients (Sect. \ref{s:DA}) into three risk groups (column 1: low, intermediate (Int), and high). Corresponding median survival in weeks and 95\% confidence intervals (CI) are given in column 2 and number of patients in each group in column 3.  Variables included in the model (columns 4--10) are Karnofsky performance score (KPS),  age, current TMZ, time since diagnosis (Dx), prior TMZ,   baseline steroid use (steroid), and  gender. }
\begin{center}
\begin{tabular}{cccccccccc}
 		&				&		& \multicolumn{7}{c}{Variables}\\		\cline{4-10}
Risk 	& Median survival	&$n$	&		&		&Current	& Time Since		& Prior	&	& \\ %
Group		&  (95\% CI)	&	 (549)	&KPS		&Age		& TMZ	& Dx				& TMZ		 &Steroid	&Gender\\ \hline

\multirow{4}{*}{Low}		&		&	\multirow{4}{*}{144}	&	$\ge 90$	&	$\le 55$	&		 &		 &	No	&		&		\\
	&	55.6 &		&	$\le 80$	&		&	No	&		&		&	No	&	Female	\\
	&	(48.6, 66.3)		&		&	$\le 80$	&		&	Yes	&	$\le 16.9$	&		&		&		 \\
	&		&		&	$\le 80$	&		&	Yes	&	$(24, 33.7]$	&		&		&		\\ \hline
\multirow{3}{*}{Int.}	&		&	\multirow{3}{*}{218}	&	$\ge 90$	&		&		&		&	 Yes	 &		&		\\
	&	32.3 	&		&	$\ge 90$	&	$>55$	&		&		&	No	&		&		\\
	&	(28.1, 36.9)	&		&	$\le 80$	&		&	Yes	&	$< 33.7$	&		&		&		 \\ \hline
\multirow{3}{*}{High}		&		&	\multirow{3}{*}{187}	&	$\le 80$	&		&	No	&		 &		 &	Yes	&		\\
	&	19.7 	&		&	$\le 80$	&		&	No	&		&		&	No	&	Male	\\
	&	(18.3, 21.9)	&		&	$\le 80$	&		&	Yes	&	$(16.9, 24]$	&		&		&		 \\ \hline

\end{tabular}
\end{center}
\label{Table: partDSA Brain}
\end{table}

\begin{table}
\caption{$L\&C_{NLL}$  stratification   of NABTC patients (Sect. \ref{s:DA}) into three risk groups (column 1: low, intermediate (Int), and high). Corresponding median survival in weeks and 95\% confidence intervals (CI) are given in column 2 and number of patients in each group in column 3.  Variables included in the model (columns 4--5) are age and  prior TMZ. }
\begin{center}
\begin{tabular}{ccccc}
Risk &	Median survival	&	$n$ 	&	\multicolumn{2}{c}{Variables	}				\\ \cline{4-5}
Group		&	 (95\% CI)	&	(549)	&	Age	&	Prior TMZ		\\  \hline	
Low	&	46.3 (40.9, 54.7)	&	176	&	$ \le 55$	&	No	\\		
Intermediate	&	27.3 (22.3, 33.3)	&	177	&	$ \le 55$	&	Yes	\\		
High	&	25.3 (21.7, 28.3)	&	196	&	$ > 55$	&		\\		 \hline
\end{tabular}
\end{center}
\label{Table: LC-CART Brain}
\end{table}

In earlier analyses of other glioma studies, \citet{Wong01081999} found
histologic diagnosis at study registration (i.e., grade), prior treatment(s), and
KPS to be associated with overall survival, whereas \citet{Carson20062007} noted initial
histology (GBM vs other; i.e., grade), age, KPS, baseline steroid use,
shorter time from original diagnosis to recurrence, and tumor outside frontal lobe to be prognostic.
As implied in earlier discussion, \cite{Wu01022010} found grade to be important in the
recursive partitioning analysis of the NCCTG trials alone and also in combination with the NABTC
trials; however, grade was not identified as being prognostic when looking only at the NABTC trials.
In combination with our own analyses of the NABTC trial data, we conclude that
prior TMZ therapy, baseline steroid use, and gender should be incorporated into the
set of variables that \cite{Wu01022010} recommend for consideration
when planning future clinical trials of high-grade glioma patients.


\section{Discussion}

The ability to successfully build a model for predicting a survival
outcome has important clinical implications.
Censored survival outcomes have inherent challenges compared to continuous and
categorical outcomes; the development of tree-based methods for
analyzing survival data deserves further attention and study. We have
introduced two methods for extending \emph{partDSA} to right-censored
outcomes: the IPCW and Brier Score weighting schemes. In simulation
studies, these two methods are observed to perform as well, and often
considerably better, than either $CART_{IPCW}$ or $L\&C_{NLL}$ using
several distinct evaluation criteria. As illustrated
in Figure \ref{Fig:MultiTT}, greater stability can be
expected from the \emph{partDSA} representation since
fewer terminal partitions are needed to
represent distinct groups of subjects
with similar levels of risk.

The $partDSA_{Brier}$ methodology holds clear promise for risk stratification
on the basis of survival probabilities, particularly so the ``integrated''
(i.e., composite loss) versions. Compared with the two other IPCW-$L_2-$loss based methods,
$partDSA_{Brier}$ maintains good performance for higher censoring levels.
In part, we attribute this good behavior
to both a sensible target of estimation and
the calculation of IPC
weights at time points bounded away from the
tail of the survivor distribution.
The $partDSA_{Brier}$ methods also make better use of the available failure time
information by including both   observations that are uncensored and
censored after a specified time cutpoint; 
whereas IPCW (for both algorithms) only directly includes uncensored
observations. 
This additional utilization of information may
reduce the variability in the estimated loss functions, resulting in
improved performance; see \cite{straw2000} for related discussion. 
Finally, the ``integrated'' (i.e., composite loss)
version of the Brier score loss function
has two
useful robustness properties: invariance to
monotone transformations of time; and,  strong
connections to estimating median, rather than mean,
survival \citep[Theorem 1]{Lawless2010}.
As a result, the performance of $partDSA_{Brier}$ 
as a risk stratification procedure
is less likely to be influenced by the presence of a
few extreme response values.

Although \emph{partDSA} has advantages over CART, one
important fixed disadvantage is its greater computational burden.  In
particular, because \emph{partDSA} iterates among three possible moves
and performs an exhaustive search of the covariate space, it will
inevitably require a significantly higher running time than CART. The
\texttt{R} package for \emph{partDSA} allows for the cross-validation
folds to be run in parallel, making runnings time feasible in
many applications \citep{Molinaro2010}, especially given that most
clinical datasets remain relatively modest in size.  In future work,
we will look further at variable selection in \emph{partDSA} and
corresponding variable importance measures.
Such extensions will serve to further improve the already-demonstrated potential
of \emph{partDSA}.



\section{Supplementary Materials}

Web Appendices, Tables, and Figures referenced in Sections
2 -- 4  are available at the end of  this paper.\vspace*{-8pt}

\section*{Acknowledgements}

We thank our colleagues at UCSF, Michael Prados and Kathleen Lamborn, for
access to the NABTC trial data as well as fruitful  discussions about the
data analysis results; and, Steve Weston, Robert Bjornson,
and Nicholas Carriero for computing assistance and advice.  This
work was supported by a CTSA Grant (UL1 RR024139 to A.M.M. and K.L.)
from the National Center for Research Resources, a component of
the National Institutes of Health (NIH), and NIH Roadmap for Medical
Research; ``Yale University Life Sciences Computing Center'' and NIH
High End Shared Instrumentation Grant [RR19895 for instrumentation];
NIH National Library of Medicine (T15 LM07056 to K.L.).\vspace*{-8pt}

\bibliographystyle{biom}
{\bibliography{refs}}
\linespread{1.6}

\label{lastpage}
\newpage

\pagestyle{empty}
\processdelayedfloats

\newpage

\pagestyle{plain}
\clearpage

\setcounter{table}{0}
\setcounter{section}{0}
\setcounter{page}{1}
\setcounter{figure}{0}
\setcounter{postfig}{0}
\setcounter{posttbl}{0}

\nomarkersintext

\linespread{1.3}
\begin{center}
  {\Large {\bf Web-based Supplemental Materials for \\[2ex]
      \textit{
        A Partitioning
  Deletion/Substitution/Addition Algorithm for Creating Survival Risk
  Groups}}} \\[2ex]

by Karen Lostritto, Robert L.\ Strawderman, and Annette M.\ Molinaro
\end{center}

\linespread{1.6}

\section*{Web Appendix A: Further detail on partDSA}\label{s:partDSAdetails}

\begin{figure}[!htb]
\center
\includegraphics[angle=0,width=7in]{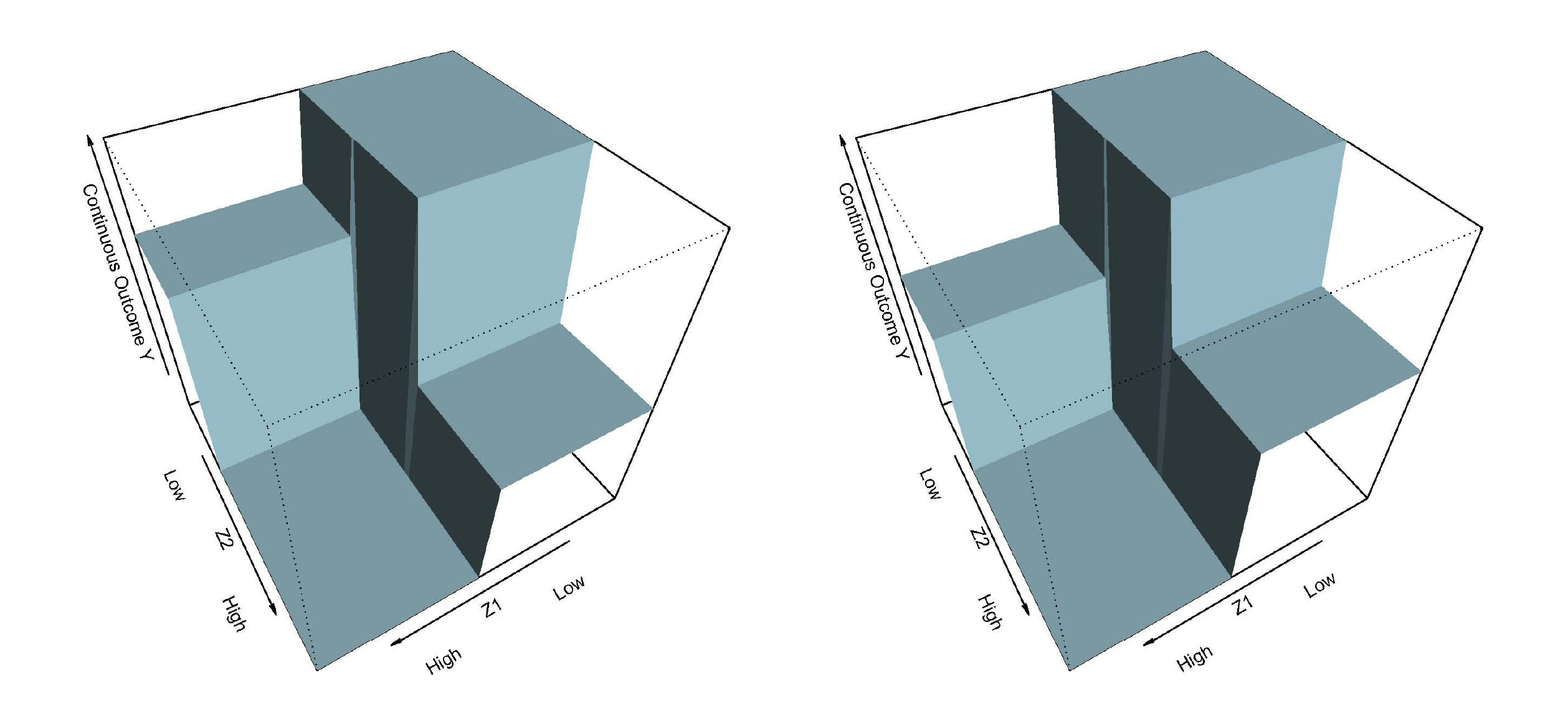}
\caption{{\em Two dimensional display of Deletion Step.} Two
disparate regions on the left form a union on the right. This union
reads: "If low on Z1  {\bf and} high on Z2
 {\bf OR} low on Z2  {\bf and} high on Z1."} \label{fig:deletion}
\end{figure}

\begin{figure}[!htbp]
\center
\includegraphics[angle=0,width=7in]{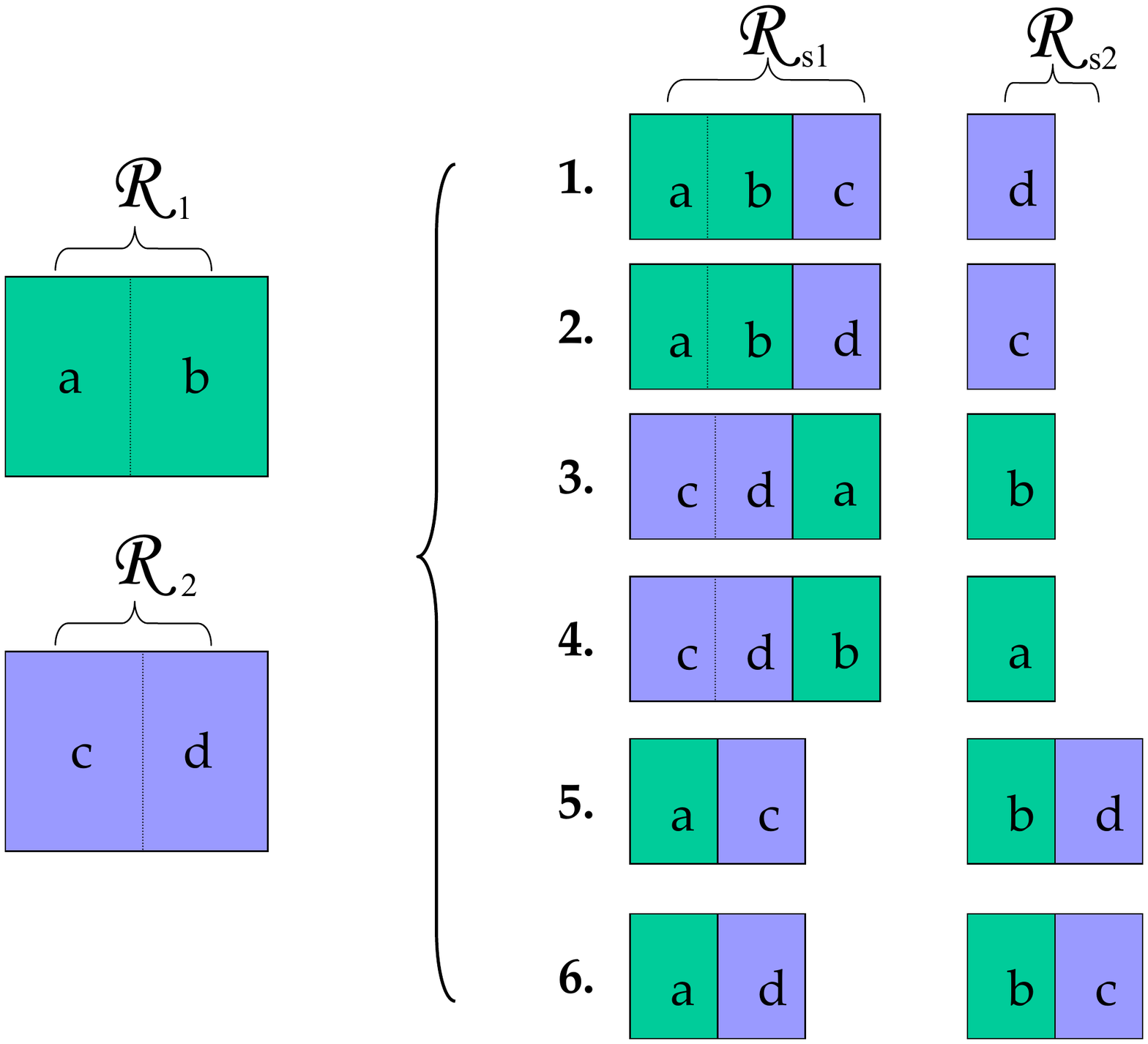}
\caption{{\em Possible Substitution Moves for two disparate
regions.} The 'best' split on Region 1 ($\mathcal{R}_1$) is found
and labeled $a$ and $b$. The 'best' split on Region 2
($\mathcal{R}_2$) is found and labeled $c$ and $d$. All (six) unique
combinations of $a,b,c, \mbox{ and } d$ are formed as $R_{s1}$ and $R_{s2}$.}
\label{fig:subset}
\end{figure}

\begin{figure}[!htbp]
\center
\includegraphics[angle=0,width=7in]{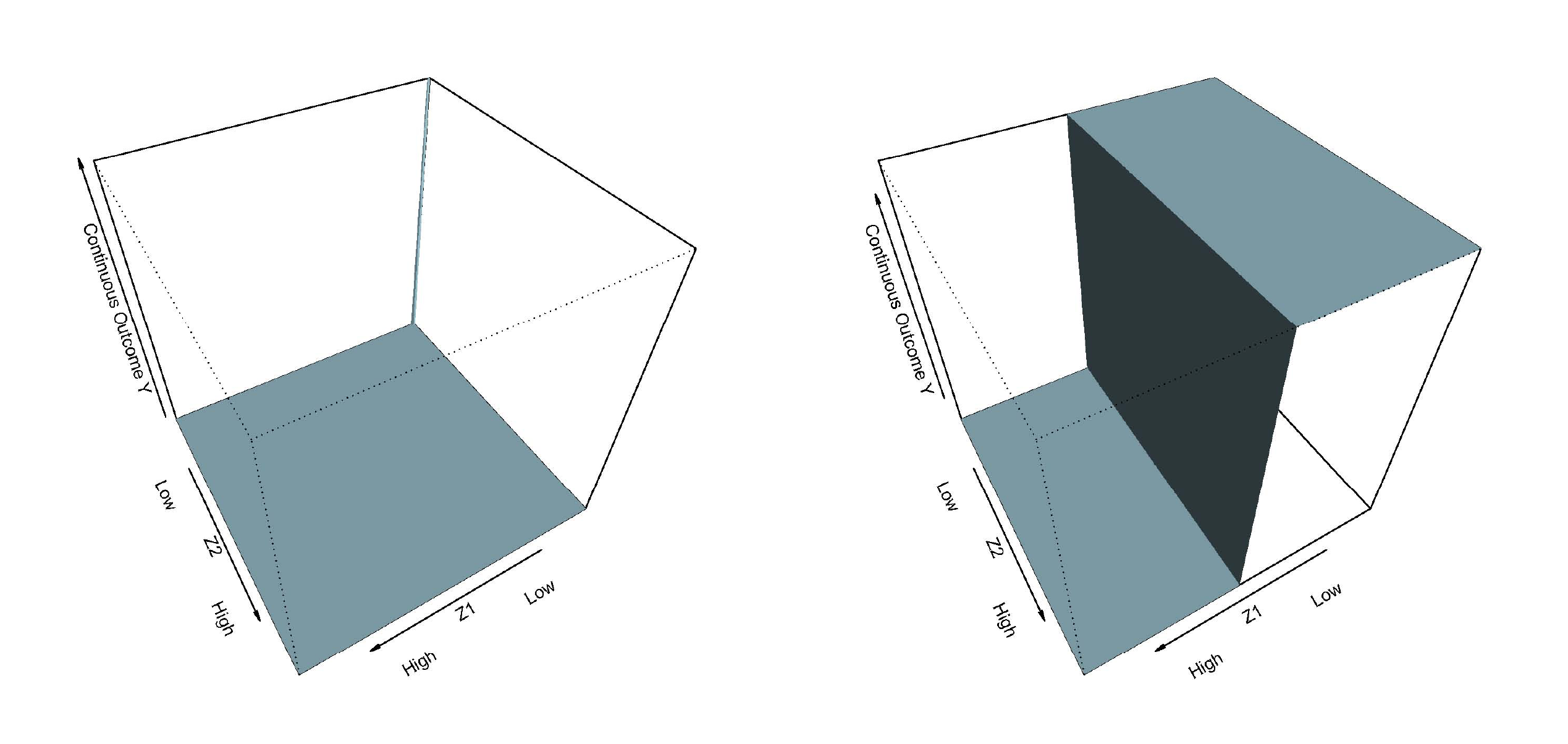}
\caption{{\em Two dimensional display of Addition Step.} The
algorithm is initiated on the left and the 'best' split is
found separating 'low' on Z1 from 'high'.} \label{fig:addition}
\end{figure}

\emph{partDSA} utilizes three moves, or step
functions, to generate  index sets (i.e., different partitionings of
the covariate space) with the goal of minimizing a risk function
over all the generated index sets. These three moves include:

\begin{itemize}
\item {\bf Deletion:} A deletion move forms a union  of two regions
of the covariate space regardless of their spatial location, i.e.,
the two regions may not be contiguous. An example is shown in Figure  \ref{fig:deletion}.

\item {\bf Substitution:} A substitution move divides two disparate regions into two subsets each and then forms combinations of the
four subsets resulting in two new regions. Thus, this step forms
unions of regions (or subsets within the regions) as well as divides
regions.  The possible subsets of two regions and combinations
thereof can be seen in Figure \ref{fig:subset}.

\item {\bf Addition:} An addition move splits one region  into
two distinct regions.  An example is shown in Figure  \ref{fig:addition}.
\end{itemize}

\processdelayedfloats
\newpage

\section*{Web Appendix B: Computational Details For Simulation Studies}
\label{WebAppend:SimDescription}

The \texttt{rpart} package  version 3.1-49 available in \texttt{R} was used for the
$L\&C_{NLL}$ method, and code from \cite{Molinaro_2004}, also based on
\texttt{rpart}, for $CART_{IPCW}$. The default tuning parameters were
used for \texttt{rpart} except as noted below.  In \emph{partDSA}, version 0.8.2 (available on R-forge),
$MB$, defined as the minimum number of observations per ``or'' statement
was set to 15.  In \texttt{rpart}, $minbuck$, the minimum number of
observations per node was set to 15.  To allow both algorithms the
opportunity to build the same size models we set \texttt{rpart}'s
$minsplit$ to 30.  The minimum percent difference ($MPD$) in
\emph{partDSA} is a threshold for the required improvement in the risk
to make a particular move (e.g., deletion) and was set to 0.05.
Additionally, both algorithms were restricted to a max of 10 nodes or
partitions. Five-fold cross validation was carried out in a standardized way
for all methods such that the partition of the observations into the five folds was
identical for each method.

As indicated in the main document, the observed survival times are truncated
to help ensure estimated censoring probabilities remain bounded away from zero.
Specifically, we set the sample-dependent truncation time $\tau$ such
that the proportion of observed follow-up times exceeding $\tau$
is 5\% of the total sample size. All follow-up times exceeding
$\tau$ are set equal to $\tau$ and are subsequently considered uncensored.
Consequently, all truncated survival times lie in $(0,\tau]$.

In the case of $\emph{partDSA}_{Brier},$ a vector of times for loss computation
must be chosen. As described in the main paper, we simply used the theoretical median
of the marginal survivor function computed under the specified model in the case of
$\emph{partDSA}_{Brier}(1 Fixed).$ In the case of $\emph{partDSA}_{Brier}(5 Even)$ and
$\emph{partDSA}_{Brier}(5 KM),$ a sample-dependent vector of 5 time points is used.
For $\emph{partDSA}_{Brier}(5 Even),$ we select the time points $(j/6)\tau$ for $j = 1,2,3,4,5$;
for $\emph{partDSA}_{Brier}(5 KM),$ we compute the Kaplan-Meier estimate of the
marginal survivor function and then use the times that respectively correspond
to survival probabilities  of 0.85, 0.7, 0.55, 0.40 and 0.25.  The
loss $BS^c(t)$ in \eqref{orig B score} is then computed for each of the 5 time points in both cases
and aggregated to obtain a composite measure of loss. In particular, if $t_1 < t_2 < t_3 < t_4 < t_5$
denote these 5 time points, we compute the weighted average loss:
\[
B^* = \sum_{i=1}^5 |\frac{t_i}{t_5}| BS^c(t_i),
\]
the weight $|t_i/t_5|$ serving to emphasize later survival differences over earlier
differences (and the absolute value allowing for the possibility of times on the log
scale); see \cite{Graf1999} for further discussion.

\processdelayedfloats

\section*{Web Appendix C: Extended Detail for Multivariate Simulation Studies}

\subsection*{C.1: Detailed Description of Evaluation Measures}

\emph{Prediction Concordance:}
To ensure comparability across the different methods of tree
construction, we define prediction concordance using
the terminal-node-specific IPCW-estimated average survival time derived
from the training set data as the predicted outcome.
For each test set, all members are classified into a terminal node for each estimated
tree based only on their covariates and assigned a corresponding estimated average
survival time as their predicted outcome. The concordance index, or c-index,
compares observed and predicted outcomes for each method in each test set by
computing the fraction of pairs in which the observation having the shorter event
time also has the shorter model-predicted event time \citep{Harrell_1982}.
Since it is possible (indeed, expected) that test subjects with different outcomes
will have tied predicted values, we report two c-indices,
one that excludes all such tied pairs ($C_p$) and another that is calculated
as the numerical average of this c-index and that calculated including
all such tied pairs ($\bar{C}_p$). The latter
typically exhibits less bias in comparison to
measures that respectively exclude and include ties
\citep{YanGreene08}.

\emph{Prediction Error:} In this method of evaluation,
two predicted outcomes are computed for each test set member for each
estimation method.  First, a predicted outcome is determined  by running all
test set subjects down the true \emph{partDSA} tree and
then again down the corresponding CART tree (cf.\ Figure \ref{Fig:MultiTT}).
The predicted outcome for subject $i$, say $\psi^{TT}_i,$ is taken to be the average
survival time of all members assigned to the same terminal node.  The second predicted
outcome is determined in exactly the same fashion, but instead runs
each test set member down the estimated tree built using the training
set; call this predicted outcome $\psi^{TS}_i.$  Comparing predictions
over all subjects measures how well the estimated tree
(i.e., structurally speaking) predicts the true subject-specific risk.
We measure the distance between these predicted outcomes using
$L_p = 1/5000 \sum_{i=1}^{5000} (\psi^{TT}_i - \psi^{TS}_i)^2.$
Evidently, $L_p = 0$ if and only if both trees classify all test set
subjects into the same risk groups, and increases away from zero as
the heterogeneity in risk group assignment, hence predicted risk, increases.

\emph{Risk Stratification:} This criterion focuses on
the ability of each method to properly separate patients into groups of
differing risk.  In particular, for each of the 1000 independent test
sets, node-specific empirical survivor functions are computed.  Then,
for each estimation method, and for the subset of the 1000 estimated
trees that consist of either two or three terminal nodes, we computed
the corresponding average survivorship and 0.025 and
0.975 percentiles on a fine grid of time points. A graphical summary of the
results is provided for each simulation study.
For \emph{partDSA}, we expect to see a high proportion of cases with only two risk
groups (i.e., survival curves) and small standard errors. For CART, we
expect to observe a high proportion of cases with three risk
groups, with two of these groups representing subpopulations
having identical survival distributions and corresponding
estimates that consequently reflect greater levels of error.

\emph{Pairwise Prediction Similarity:} This last criterion
also examines the separation of patients into risk groups, targeting
the ability of each method to identify groups of subjects having a
similar level of risk. Given the underlying tree model,  the
terminal node (hence risk group) for each observation is known. Define
$I_T(i,j)$ to be 1 if observations $i$ and $j$ are grouped into the same
terminal node (i.e., into the same risk group) under
this true model and set it equal to 0 otherwise. Then,
$I_T(i,j)$ is known for each of the ${n \choose 2}$
unique pairs in each test set. For structures estimated by
\emph{partDSA} and CART, we can compute
analogous measures of agreement/discrepancy for all
unique test set pairs; generically, let this measure
be denoted by $I_M(i,j).$ Then, $|I_T(i,j)-I_M(i,j)| > 0$ if and only if an
estimated tree incorrectly groups observations $i$ and $j.$
Similarly to \cite{Chip01}, define the distance measure
\begin{equation}
D_p = 1- \sum_{i=1}^{n} \sum_{j>i} \frac{ | I_T(i,j)-I_M(i,j) | }{{n \choose 2}}
\end{equation}
With perfect agreement, $D_p = 1$; with perfect disagreement, $D_p = 0$.
We report the average value of $D_p$ computed over all 1000 test sets.
An advantage of this metric is its independence from both tree topology and
the actual predictions associated with each terminal node.

\subsection*{C.2: High Signal, Covariate-Dependent Censoring}

\begin{figure}
\centering
\includegraphics[width=6in,height=6in]{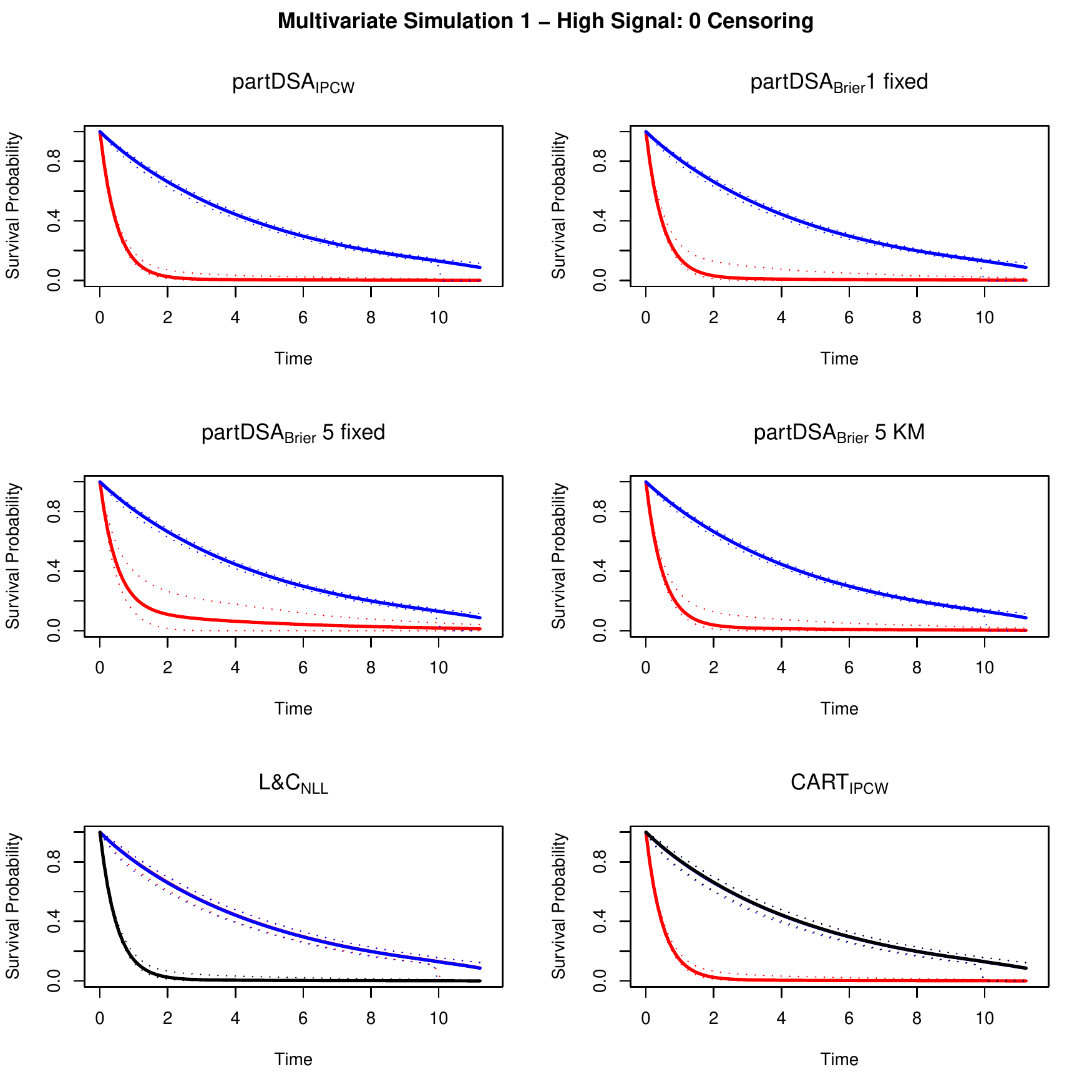}
\caption{High Signal, Covariate-Dependent Censoring: Kaplan-Meier plots for
six methods to illustrate the survival
  experience of chosen risk groups (Section \ref{s:MVsims1}, 0\% censoring).}
\label{WebFig:KMMulti10}
\end{figure}

\begin{figure}
\begin{center}
\includegraphics[width=6in]{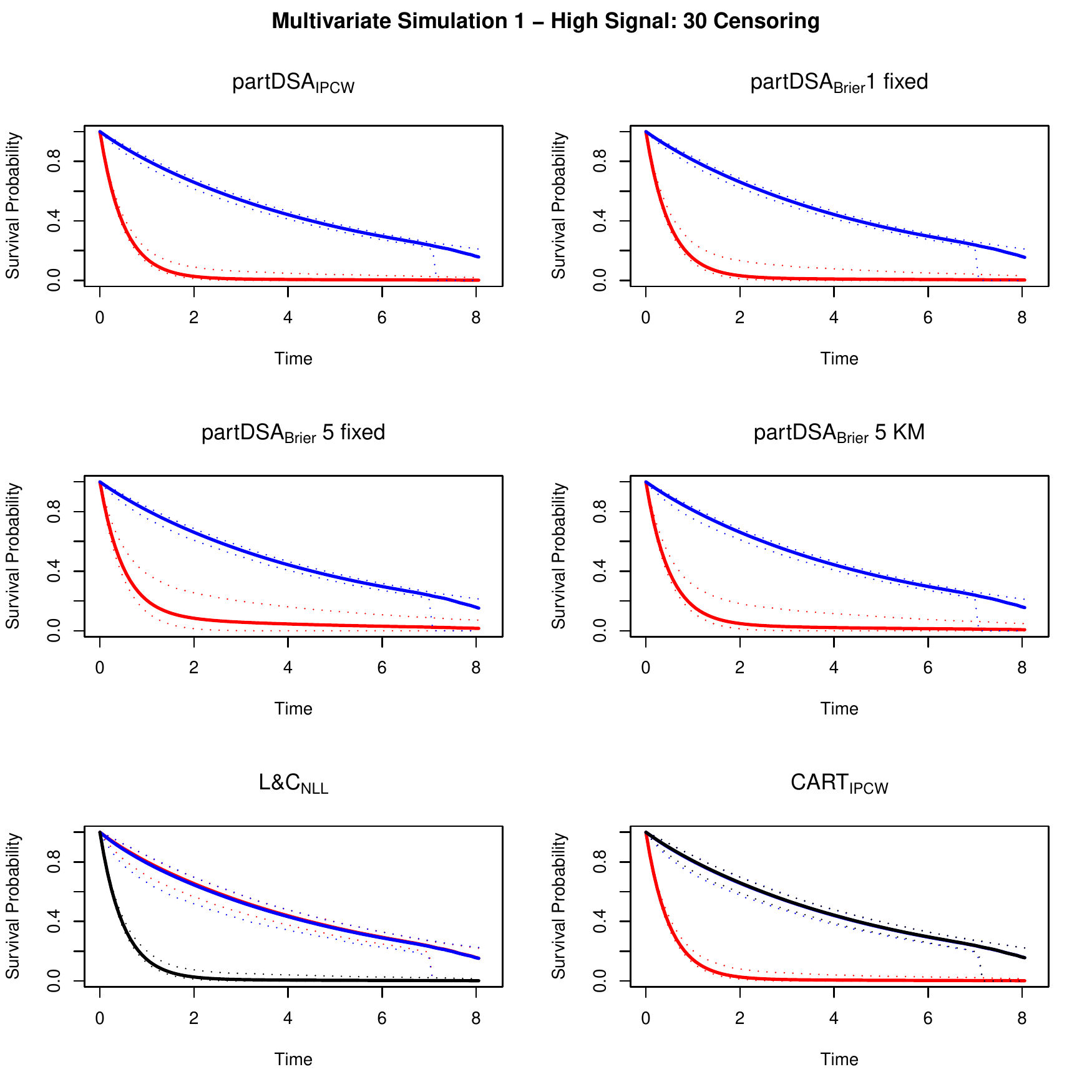}
\caption{High Signal, Covariate-Dependent Censoring: Kaplan-Meier plots for six methods to illustrate the survival
  experience of chosen risk groups (Section \ref{s:MVsims1}, 30\% censoring).}

\label{Fig:KMMulti1}
\end{center}
\end{figure}

\begin{figure*}
\begin{center}
\includegraphics[width=6in]{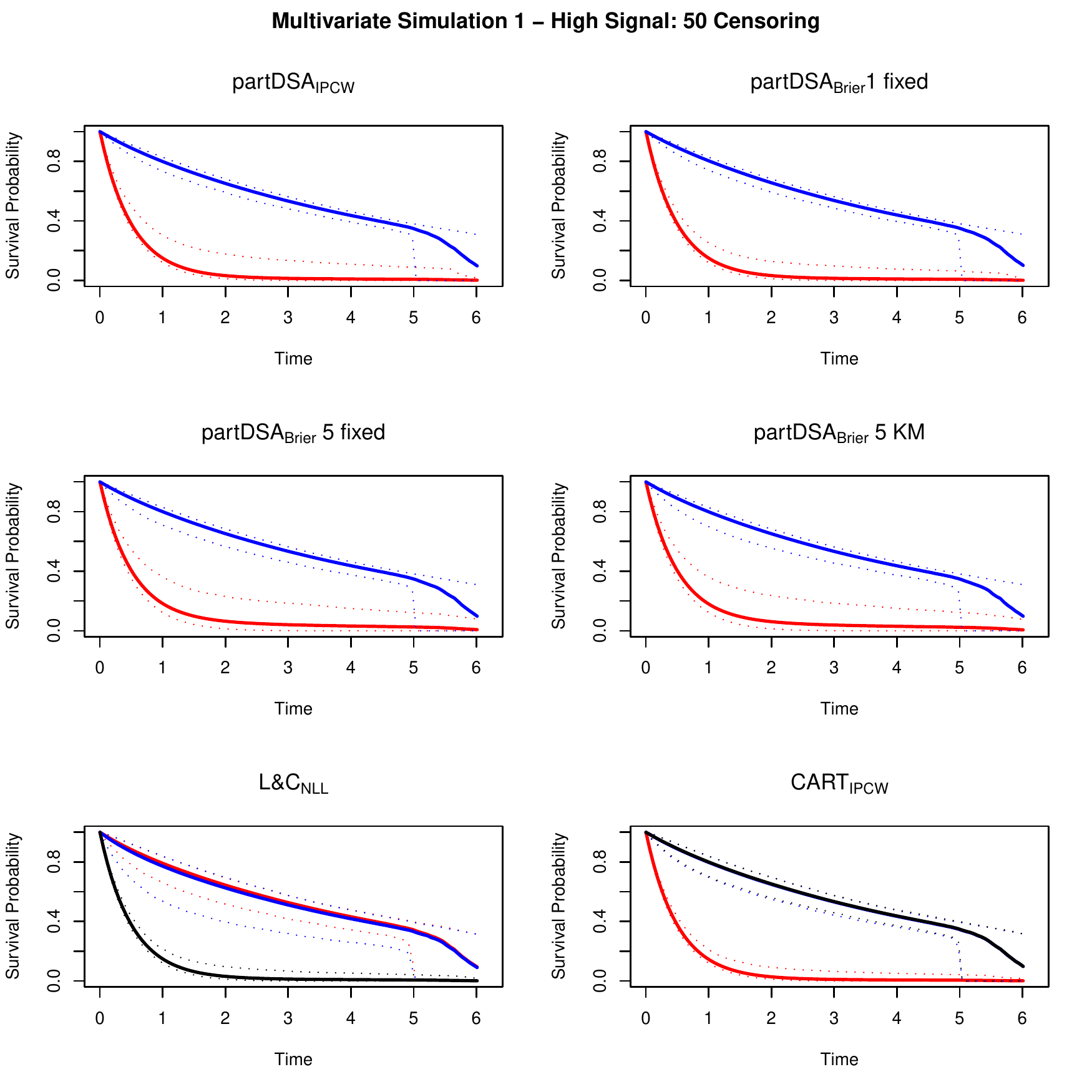}
\caption{High Signal, Covariate-Dependent Censoring: Kaplan-Meier plots for six methods to
illustrate the survival experience of chosen risk groups  (Section \ref{s:MVsims1}, 50\% censoring).}
\label{WebFig:KMMulti150}
\end{center}
\end{figure*}

\begin{table}
 \caption{High Signal, Covariate-Dependent Setting: proportion of the 1000 models
 at each given size for each method at the three censoring levels.\label{WebTable:MV1_1ModelSize}}
 \begin{center}
 \begin{tabular}{lrrrr}\hline\hline
\multicolumn{1}{l}{}&\multicolumn{1}{c}{Root Node}&\multicolumn{1}{c}{2}&\multicolumn{1}{c}{3}&\multicolumn{1}{c}{4+}\tabularnewline
\hline
{\bfseries 0Cens}&&&&\tabularnewline
~~$L\&C_{NLL}$&$0.001$&$0.001$&$0.869$&$0.129$\tabularnewline
~~$CART_{IPCW}$&$0.018$&$0.001$&$0.861$&$0.120$\tabularnewline
~~$\emph{partDSA}_{IPCW}$&$$&$0.862$&$0.117$&$0.021$\tabularnewline
~~$\emph{partDSA}_{Brier} 1 fixed$&$$&$0.821$&$0.140$&$0.039$\tabularnewline
~~$\emph{partDSA}_{Brier} 5 even$&$0.017$&$0.625$&$0.281$&$0.077$\tabularnewline
~~$\emph{partDSA}_{Brier} 5 KM$&$$&$0.830$&$0.148$&$0.022$\tabularnewline
\hline
{\bfseries 30Cens}&&&&\tabularnewline
~~$L\&C_{NLL}$&$0.004$&$0.003$&$0.829$&$0.164$\tabularnewline
~~$CART_{IPCW}$&$0.047$&$0.002$&$0.823$&$0.128$\tabularnewline
~~$\emph{partDSA}_{IPCW}$&$$&$0.814$&$0.151$&$0.035$\tabularnewline
~~$\emph{partDSA}_{Brier} 1 fixed$&$$&$0.834$&$0.127$&$0.039$\tabularnewline
~~$\emph{partDSA}_{Brier} 5 even$&$0.014$&$0.707$&$0.195$&$0.084$\tabularnewline
~~$\emph{partDSA}_{Brier} 5 KM$&$$&$0.817$&$0.143$&$0.040$\tabularnewline
\hline
{\bfseries 50Cens}&&&&\tabularnewline
~~$L\&C_{NLL}$&$0.043$&$0.029$&$0.783$&$0.145$\tabularnewline
~~$CART_{IPCW}$&$0.094$&$0.010$&$0.785$&$0.111$\tabularnewline
~~$\emph{partDSA}_{IPCW}$&$0.001$&$0.728$&$0.205$&$0.066$\tabularnewline
~~$\emph{partDSA}_{Brier} 1 fixed$&$0.001$&$0.813$&$0.146$&$0.040$\tabularnewline
~~$\emph{partDSA}_{Brier} 5 even$&$0.035$&$0.694$&$0.183$&$0.088$\tabularnewline
~~$\emph{partDSA}_{Brier} 5 KM$&$0.021$&$0.702$&$0.186$&$0.091$\tabularnewline
\hline
\end{tabular}
\end{center}
\end{table}


\processdelayedfloats
\newpage

\subsection*{C.3: Low Signal, Covariate-Dependent Censoring}

%
\begin{figure*}
\begin{center}
\includegraphics[width=6in]{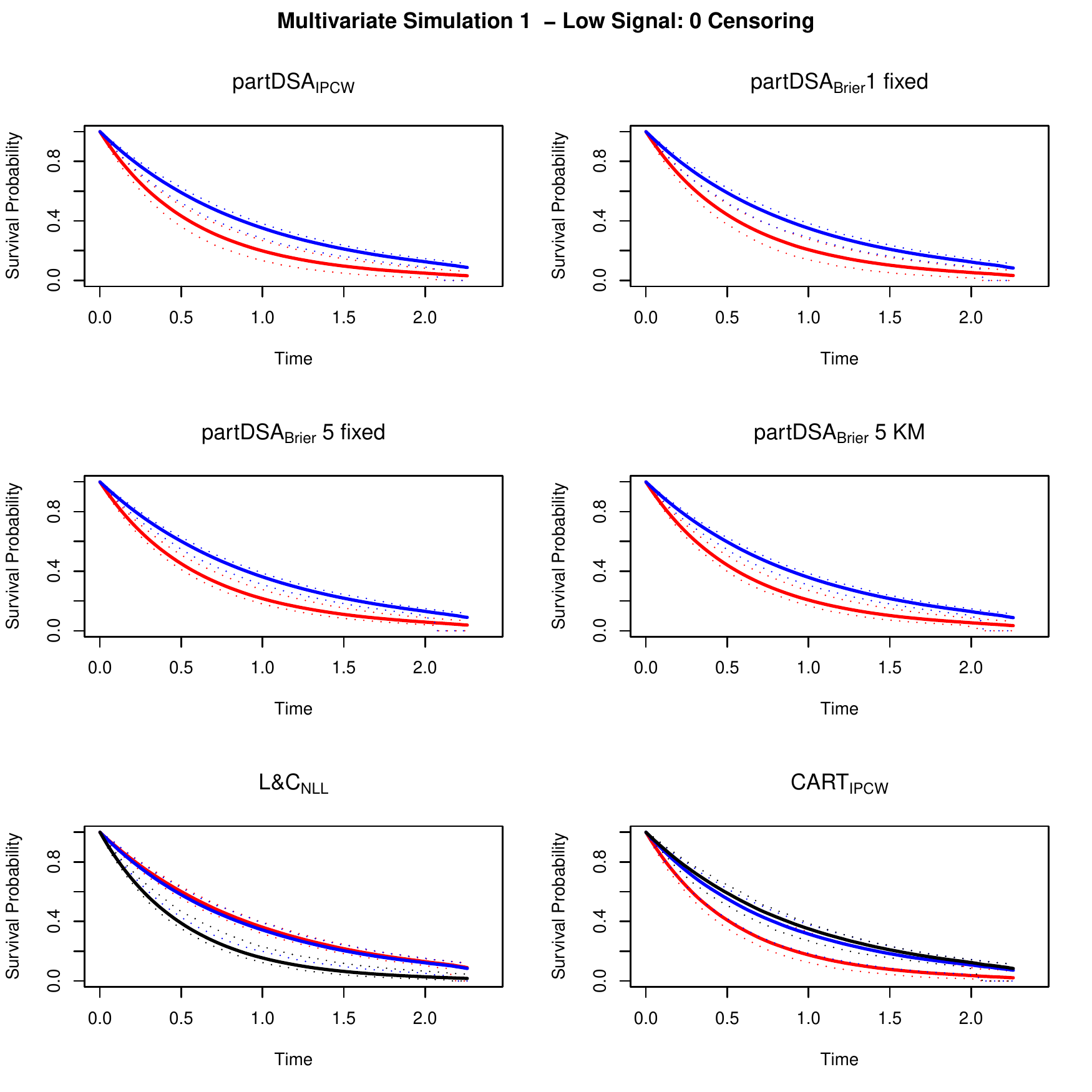}
\caption{Low Signal, Covariate-Dependent Censoring: Kaplan-Meier plots for six methods to
  illustrate the survival
  experience of chosen risk groups (Section \ref{s:MVsims1}, 0\% censoring).}
\label{WebFig:KMMulti1Low0}
\end{center}
\end{figure*}

\begin{figure}
\begin{center}
\includegraphics[width=6in]{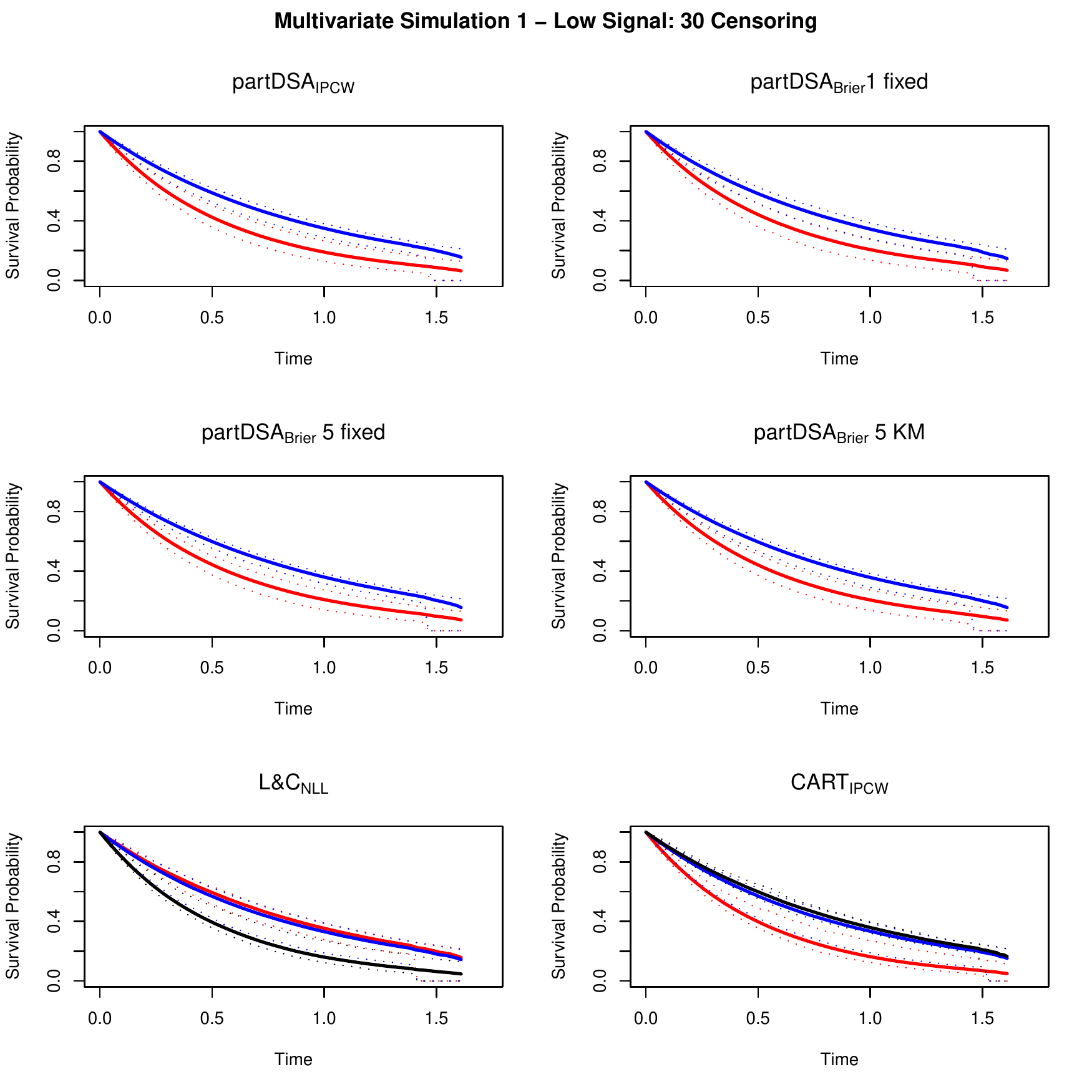}
\caption{Low Signal, Covariate-Dependent Censoring: Kaplan-Meier plots for six methods to illustrate the survival
  experience of chosen risk groups (Section \ref{s:MVsims1}, 30\% censoring).}
\label{Fig:KMMulti1Low}
\end{center}
\end{figure}

\begin{figure*}
\begin{center}
\includegraphics[width=6in]{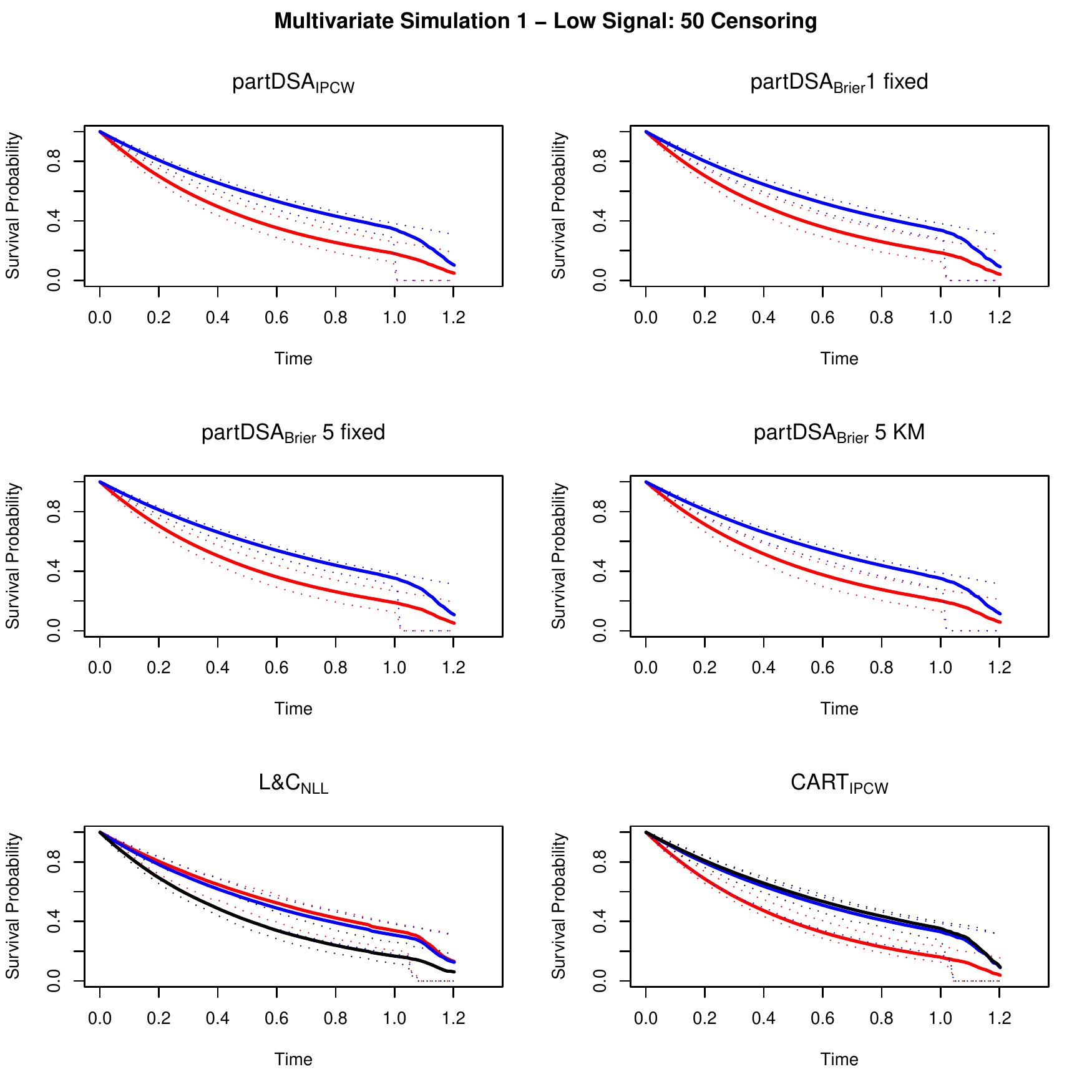}
\caption{Low Signal, Covariate-Dependent Censoring: Kaplan-Meier plots for six methods to illustrate the survival experience of chosen risk groups  (Section \ref{s:MVsims1}, 50\%
  censoring).}
\label{WebFig:KMMulti1Low50}
\end{center}
\end{figure*}

\begin{table}
 \caption{Low Signal, Covariate-Dependent Setting:
Proportion of the 1000 models at each given size for each method at the three censoring levels.
\label{WebTable:MV1_2ModelSize}}
 \begin{center}
 \begin{tabular}{lrrrr}\hline\hline
\multicolumn{1}{l}{}&\multicolumn{1}{c}{Root Node}&\multicolumn{1}{c}{2}&\multicolumn{1}{c}{3}&\multicolumn{1}{c}{4+}\tabularnewline
\hline
{\bfseries 0Cens}&&&&\tabularnewline
~~$L\&C_{NLL}$&$0.458$&$0.287$&$0.207$&$0.048$\tabularnewline
~~$CART_{IPCW}$&$0.559$&$0.320$&$0.096$&$0.025$\tabularnewline
~~$\emph{partDSA}_{IPCW} $&$0.635$&$0.290$&$0.062$&$0.013$\tabularnewline
~~$\emph{partDSA}_{Brier} 1 fixed$&$0.693$&$0.239$&$0.055$&$0.013$\tabularnewline
~~$\emph{partDSA}_{Brier} 5 even$&$0.584$&$0.343$&$0.062$&$0.011$\tabularnewline
~~$\emph{partDSA}_{Brier} 5 KM$&$0.522$&$0.336$&$0.121$&$0.021$\tabularnewline
\hline
{\bfseries 30Cens}&&&&\tabularnewline
~~$L\&C_{NLL}$&$0.662$&$0.181$&$0.120$&$0.037$\tabularnewline
~~$CART_{IPCW}$&$0.587$&$0.281$&$0.112$&$0.020$\tabularnewline
~~$\emph{partDSA}_{IPCW}$&$0.645$&$0.272$&$0.068$&$0.015$\tabularnewline
~~$\emph{partDSA}_{Brier} 1 fixed$&$0.778$&$0.171$&$0.038$&$0.013$\tabularnewline
~~$\emph{partDSA}_{Brier} 5 even$&$0.621$&$0.271$&$0.077$&$0.031$\tabularnewline
~~$\emph{partDSA}_{Brier} 5 KM$&$0.580$&$0.313$&$0.089$&$0.018$\tabularnewline
\hline
{\bfseries 50Cens}&&&&\tabularnewline
~~$L\&C_{NLL}$&$0.815$&$0.115$&$0.055$&$0.015$\tabularnewline
~~$CART_{IPCW}$&$0.597$&$0.263$&$0.114$&$0.026$\tabularnewline
~~$\emph{partDSA}_{IPCW}$&$0.549$&$0.351$&$0.072$&$0.028$\tabularnewline
~~$\emph{partDSA}_{Brier} 1 fixed$&$0.780$&$0.159$&$0.043$&$0.018$\tabularnewline
~~$\emph{partDSA}_{Brier} 5 even$&$0.453$&$0.386$&$0.112$&$0.049$\tabularnewline
~~$\emph{partDSA}_{Brier} 5 KM$&$0.557$&$0.309$&$0.089$&$0.045$\tabularnewline

\hline
\end{tabular}
\end{center}
\end{table}

\processdelayedfloats
\newpage


\subsection*{C.4: High Signal, Covariate-Independent Censoring}

\begin{figure*}
\begin{center}
\includegraphics[width=6in]{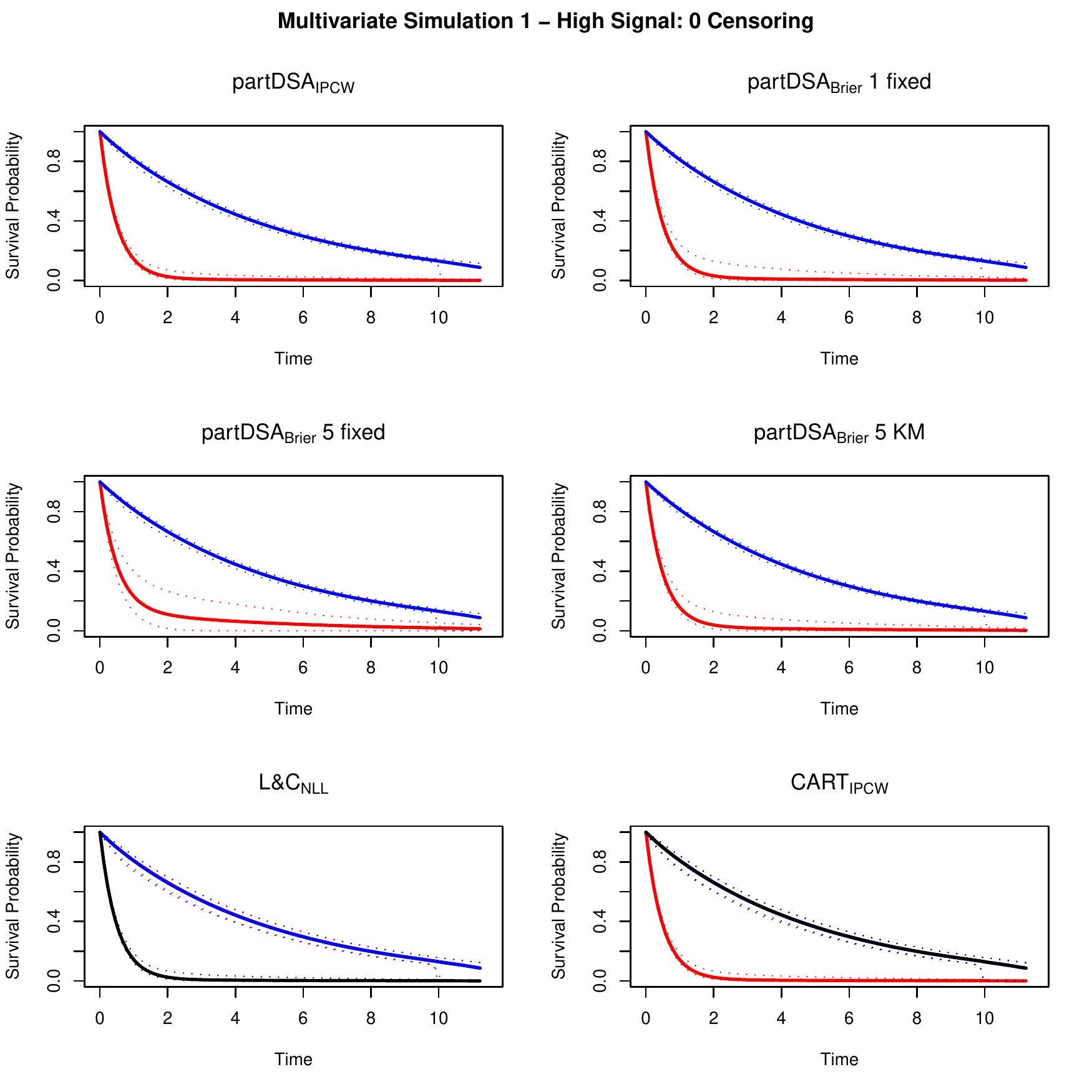}
\caption{High Signal, Covariate-Independent Censoring:
Kaplan-Meier plots for six methods to illustrate the survival
  experience of chosen risk groups (Section \ref{s:MVsims1}, 0\% censoring).}
\label{WebFig:KMMulti1nonInf0}
\end{center}
\end{figure*}

\begin{figure*}
\begin{center}
\includegraphics[width=6in]{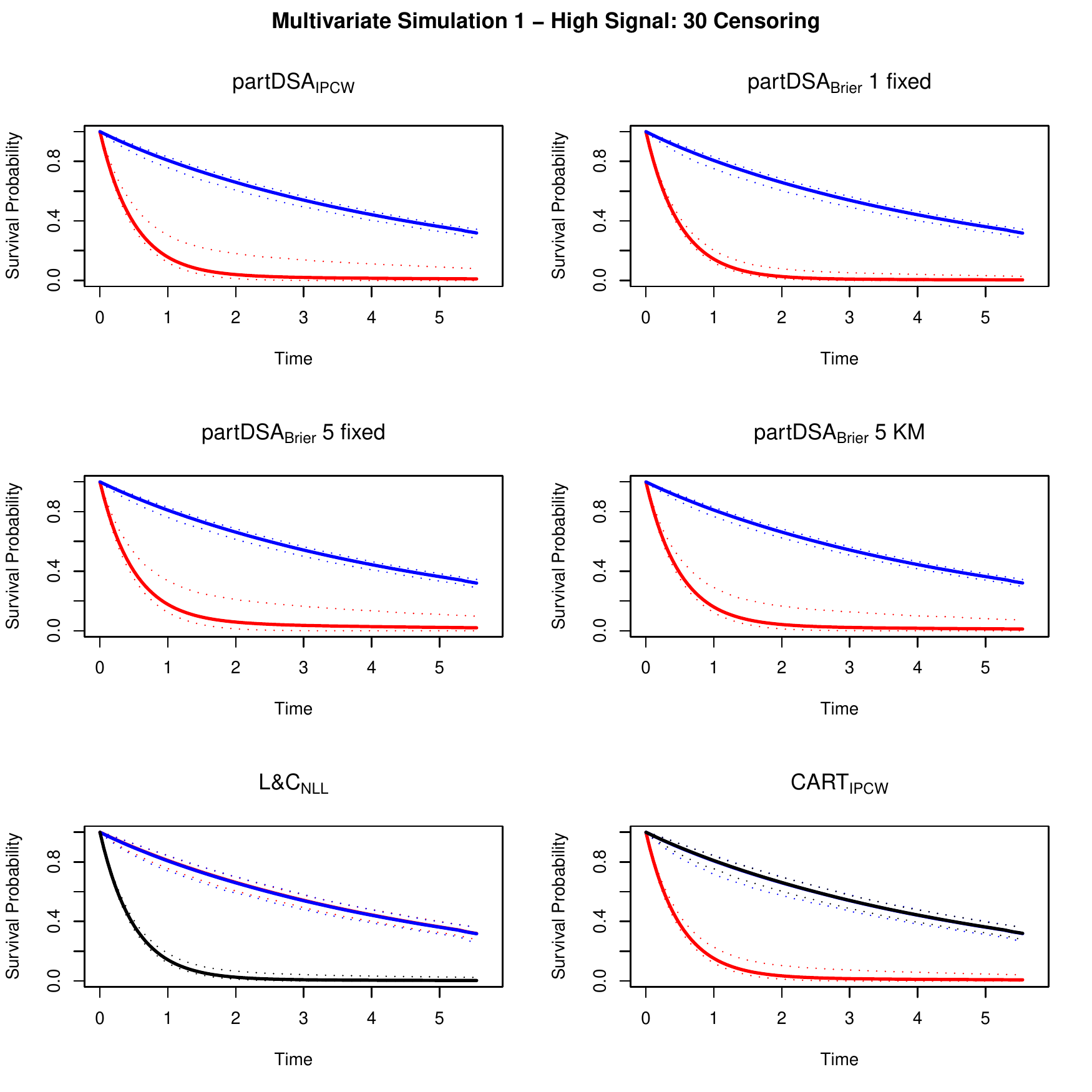}
\caption{High Signal, Covariate-Independent Censoring: Kaplan-Meier plots for six methods to illustrate the survival
  experience of chosen risk groups (Section \ref{s:MVsims1}, 30\% censoring).}
\label{WebFig:KMMulti1nonInf30}
\end{center}
\end{figure*}

\begin{figure*}
\begin{center}
\includegraphics[width=6in]{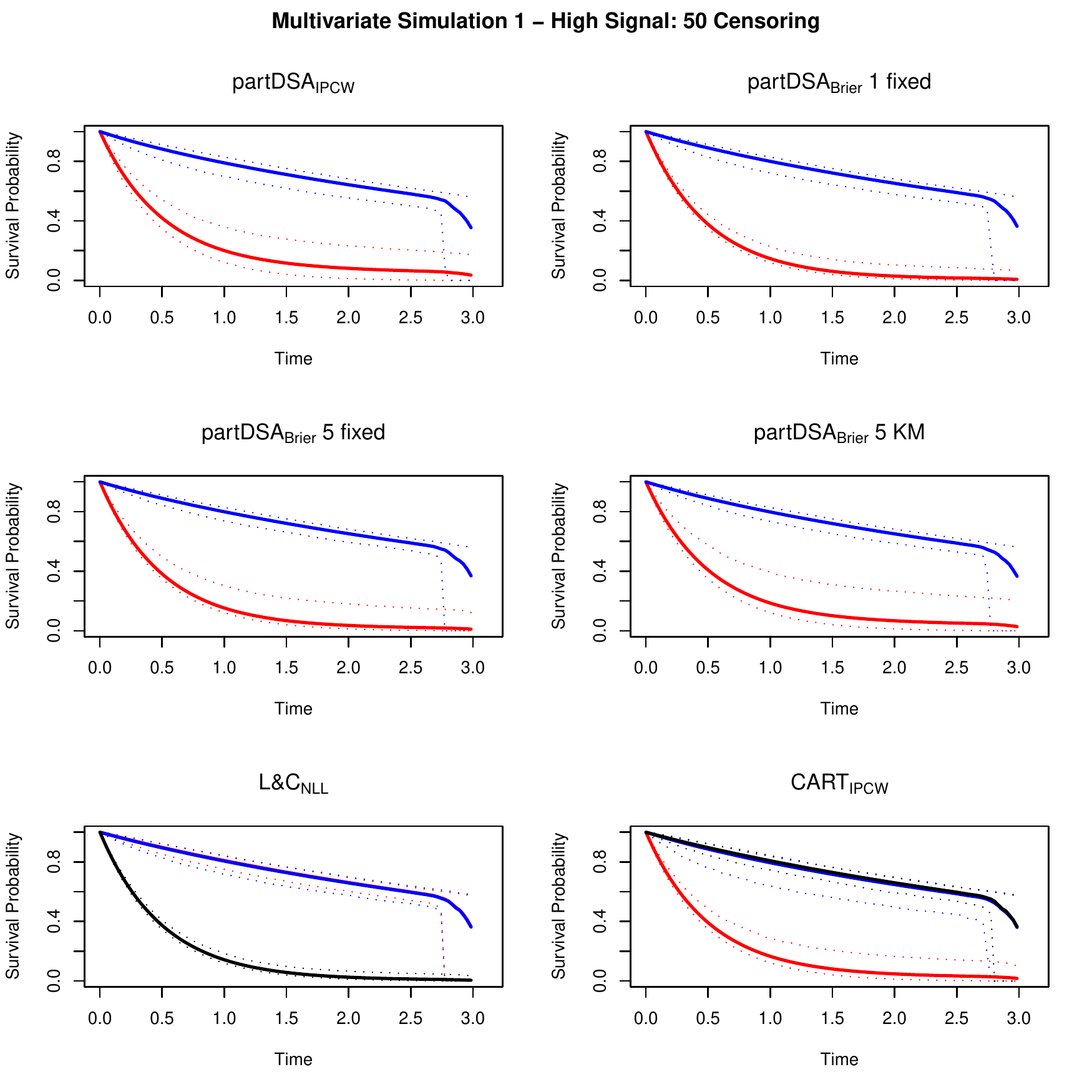}
\caption{High Signal, Covariate-Independent Censoring: Kaplan-Meier plots for six methods to illustrate the survival experience of chosen risk groups  (Section \ref{s:MVsims1}, 50\% censoring).}
\label{WebFig:KMMulti1nonInf50}
\end{center}
\end{figure*}


\begin{table}
  \caption{High Signal, Covariate-Independent Censoring: see caption of Table
  \ref{Table:MultiSim1_1} for details.} 
\begin{center}
\begin{tabular}{ c  c | c  c  c | c  c | c} 

\multicolumn{2}{c}{    }  &   \multicolumn{3}{c}{$\emph{partDSA}_{Brier}$}  &     \multicolumn{2}{c}{IPCW}     & \\ \hline
Censoring &Criteria & $1 Fixed$ & $5 Even$ & $5 KM$ & $\emph{partDSA}$ & $CART$ & $L\&C_{NLL}$ \\  \hline 
\hline 
& True Model Size & 2.00 & 2.00 & 2.00 & 2.00 & 3.00 & 3.00 \\ \hline
\multirow{4}{*}{$0\%$/$0\%$}
&Fitted Size&2.198&2.450&2.228&2.170&3.106&3.192\\
&\# Predictors&2.135&2.095&2.195&2.111&2.047&2.076\\
&\# W1-W2&1.998&1.809&2.000&1.999&1.962&1.996\\
&\# W3-W5&0.137&0.286&0.195&0.112&0.085&0.080\\
&$C_p$&0.879&0.828&0.88&0.891&0.812&0.809\\
&$\bar{C}_p$&0.782&0.748&0.783&0.789&0.748&0.747\\
&$L_p$ &0.082&0.27&0.076&0.061&0.076&0.072\\
&$D_p$ &0.944&0.838&0.946&0.964&0.845&0.842\\
\hline
\multirow{4}{*}{$30\%$/$26.1\%$}
&Fitted Size&2.332&2.416&2.213&2.344&3.003&3.135\\
&\# Predictors&2.208&2.257&2.142&2.213&1.922&2.054\\
&\# W1-W2&1.986&1.958&1.999&1.985&1.839&1.999\\
&\# W3-W5&0.222&0.299&0.143&0.228&0.083&0.055\\
&$C_p$&0.866&0.853&0.884&0.869&0.811&0.816\\
&$\bar{C}_p$&0.777&0.768&0.787&0.778&0.743&0.754\\
&$L_p$ &0.105&0.152&0.078&0.105&0.162&0.058\\
&$D_p$ &0.913&0.884&0.948&0.921&0.813&0.848\\
\hline
\multirow{4}{*}{$50\%$/$44.9\%$}
&Fitted Size&2.576&2.480&2.292&2.366&2.672&3.091\\
&\# Predictors&2.225&2.266&2.195&2.034&1.619&2.038\\
&\# W1-W2&1.875&1.982&1.990&1.753&1.551&1.997\\
&\# W3-W5&0.350&0.284&0.205&0.281&0.068&0.041\\
&$C_p$&0.842&0.858&0.879&0.844&0.82&0.833\\
&$\bar{C}_p$&0.766&0.78&0.791&0.763&0.746&0.772\\
&$L_p$ &0.159&0.096&0.082&0.212&0.235&0.043\\
&$D_p$ &0.837&0.884&0.928&0.84&0.773&0.85\\
\hline
\end{tabular}
\end{center}
\label{WebTable:MultiSim1_1nonInf}
\end{table}

\begin{table}
 \caption{High Signal, Covariate-Independent Setting: Proportion of the 1000 models at each given size for each method at the three censoring levels.\label{WebTable:MV1_1nonInfModelSize}}
 \begin{center}
 \begin{tabular}{lrrrr}\hline\hline
\multicolumn{1}{l}{}&\multicolumn{1}{c}{Root Node}&\multicolumn{1}{c}{2}&\multicolumn{1}{c}{3}&\multicolumn{1}{c}{4+}\tabularnewline
\hline
{\bfseries 0Cens}&&&&\tabularnewline
~~$L\&C_{NLL}$&$0.001$&$0.001$&$0.869$&$0.129$\tabularnewline
~~$CART_{IPCW}$&$0.018$&$0.001$&$0.861$&$0.120$\tabularnewline
~~$\emph{partDSA}_{IPCW}$&$$&$0.862$&$0.117$&$0.021$\tabularnewline
~~$\emph{partDSA}_{Brier} 1 fixed$&$$&$0.821$&$0.140$&$0.039$\tabularnewline
~~$\emph{partDSA}_{Brier} 5 even$&$0.017$&$0.625$&$0.281$&$0.077$\tabularnewline
~~$\emph{partDSA}_{Brier} 5 KM$&$$&$0.830$&$0.148$&$0.022$\tabularnewline
\hline
{\bfseries 30Cens}&&&&\tabularnewline
~~$L\&C_{NLL}$&$$&$$&$0.915$&$0.085$\tabularnewline
~~$CART_{IPCW}$&$0.070$&$0.021$&$0.774$&$0.135$\tabularnewline
~~$\emph{partDSA}_{IPCW}$&$$&$0.748$&$0.188$&$0.064$\tabularnewline
~~$\emph{partDSA}_{Brier} 1 fixed$&$$&$0.827$&$0.134$&$0.039$\tabularnewline
~~$\emph{partDSA}_{Brier} 5 even$&$0.008$&$0.720$&$0.178$&$0.094$\tabularnewline
~~$\emph{partDSA}_{Brier} 5 KM$&$0.004$&$0.754$&$0.178$&$0.064$\tabularnewline
\hline
{\bfseries 50Cens}&&&&\tabularnewline
~~$L\&C_{NLL}$&$0.001$&$0.001$&$0.930$&$0.068$\tabularnewline
~~$CART_{IPCW}$&$0.107$&$0.227$&$0.567$&$0.099$\tabularnewline
~~$\emph{partDSA}_{IPCW}$&$0.026$&$0.691$&$0.210$&$0.073$\tabularnewline
~~$\emph{partDSA}_{Brier} 1 fixed$&$0.002$&$0.781$&$0.160$&$0.057$\tabularnewline
~~$\emph{partDSA}_{Brier} 5 even$&$$&$0.645$&$0.267$&$0.088$\tabularnewline
~~$\emph{partDSA}_{Brier} 5 KM$&$0.022$&$0.595$&$0.243$&$0.140$\tabularnewline
\hline
\end{tabular}
\end{center}
\end{table}


\processdelayedfloats
\newpage

\subsection*{C.5: Low Signal, Covariate-Independent Censoring}

\begin{figure*}
\begin{center}
\includegraphics[width=6in]{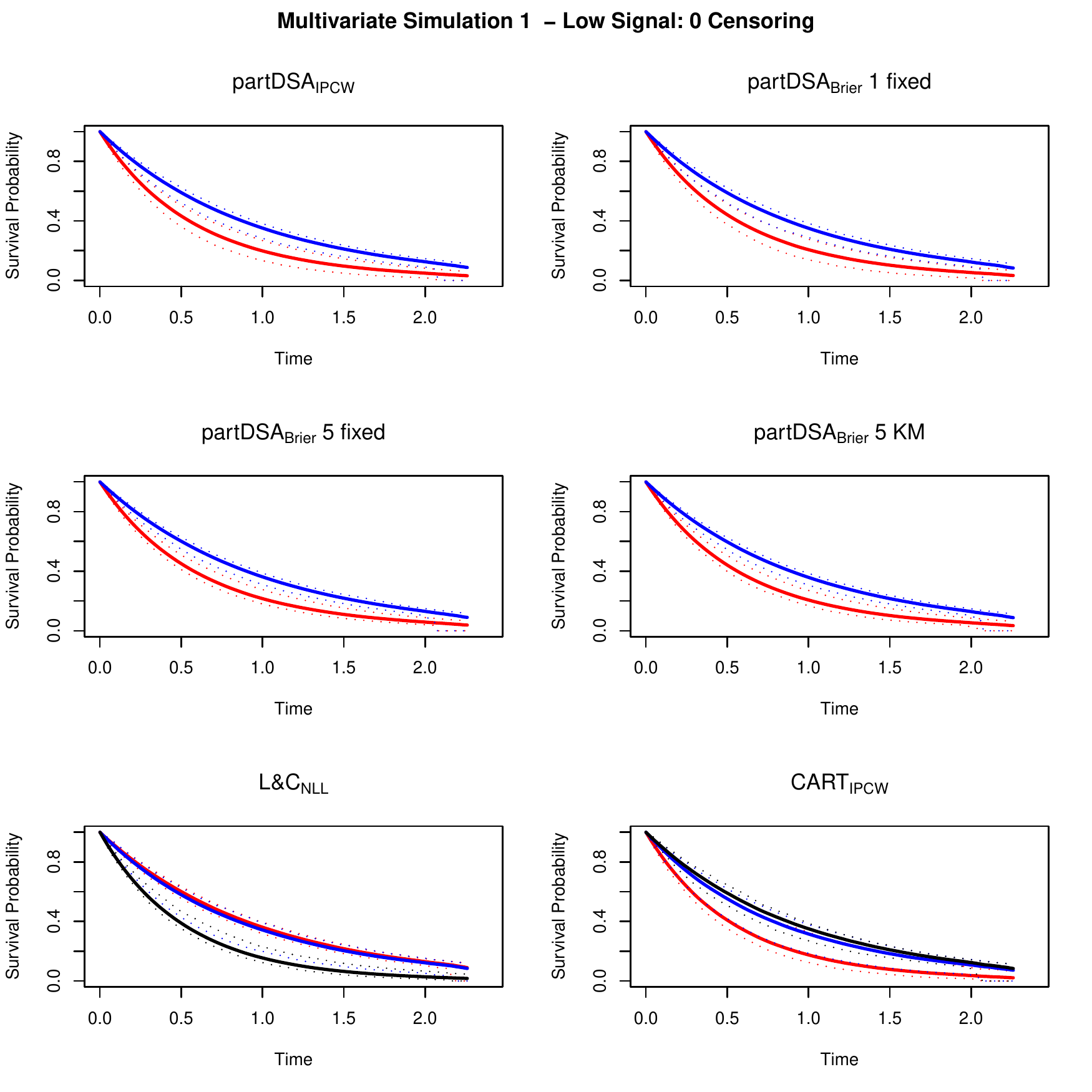}
\caption{Low Signal, Covariate-Independent Censoring: Kaplan-Meier plots for six methods to illustrate the survival
  experience of chosen risk groups (Section \ref{s:MVsims1}, 0\% censoring).}
\label{WebFig:KMMulti1LownonInf0}
\end{center}
\end{figure*}

\begin{figure*}
\begin{center}
\includegraphics[width=6in]{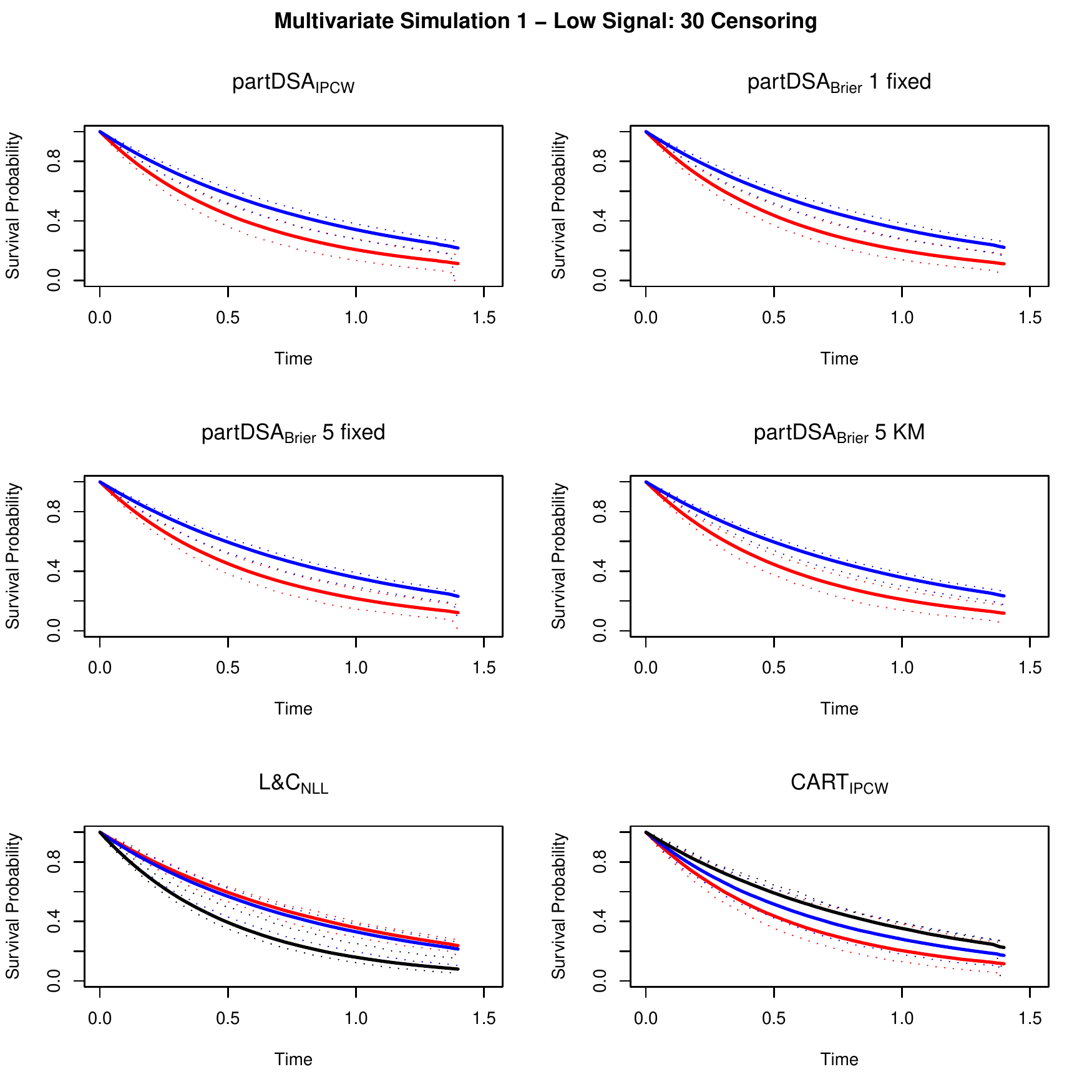}
\caption{Low Signal, Covariate-Independent Censoring: Kaplan-Meier plots for six methods to illustrate the survival
  experience of chosen risk groups (Section \ref{s:MVsims1}, 30\% censoring).}
\label{WebFig:KMMulti1LownonInf30}
\end{center}
\end{figure*}

\begin{figure*}
\begin{center}
\includegraphics[width=6in]{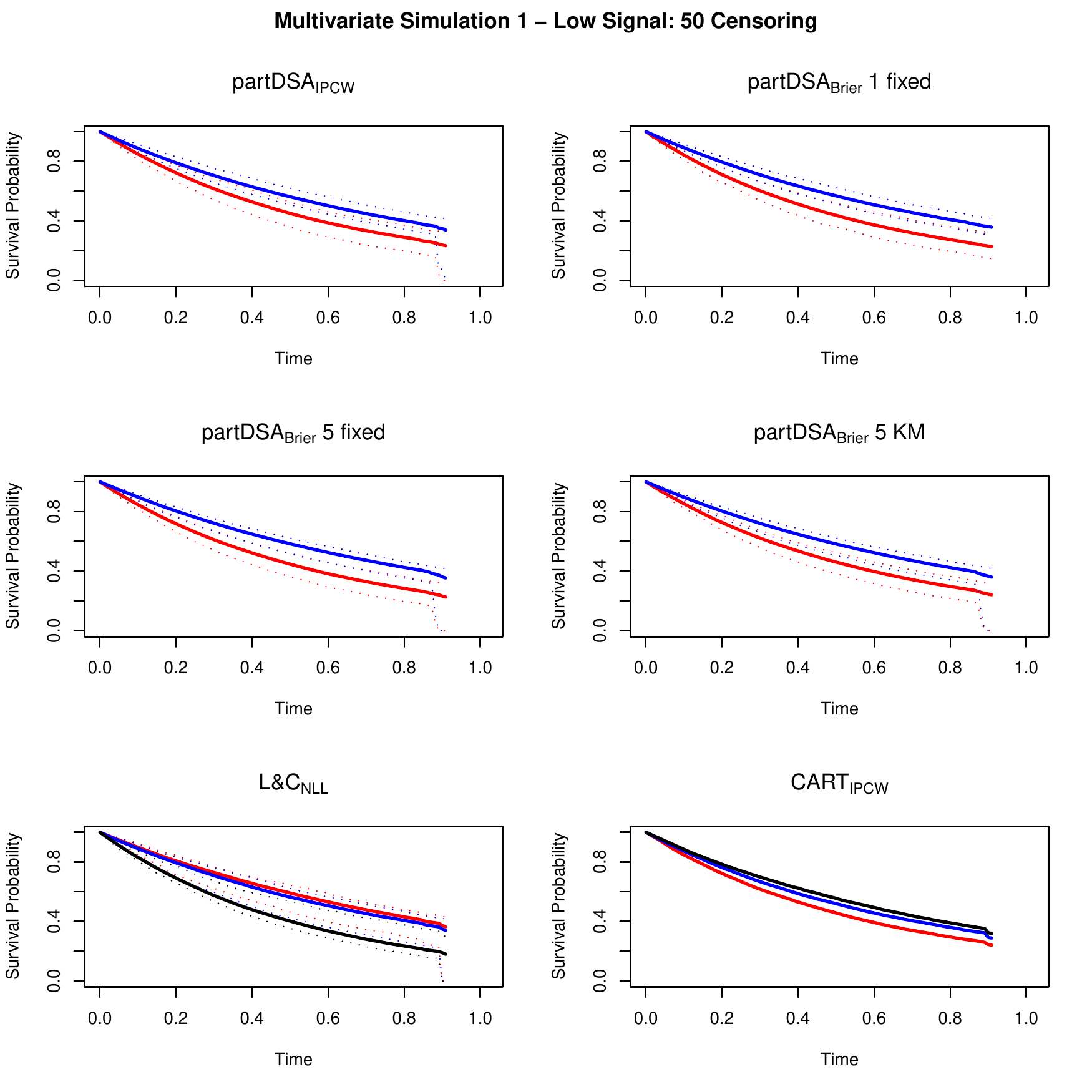}
\caption{Low Signal, Covariate-Independent Censoring: Kaplan-Meier plots for six methods to illustrate the survival experience of chosen risk groups  (Section \ref{s:MVsims1}, 50\% censoring).}
\label{WebFig:KMMulti1LownonInf50}
\end{center}
\end{figure*}

\begin{table}
  \caption{Low Signal, Covariate-Independent Censoring: see caption of Table
  \ref{Table:MultiSim1_2} for details.}
\begin{center}
\begin{tabular}{ c  c | c  c  c | c  c | c} 

\multicolumn{2}{c}{    }  &   \multicolumn{3}{c}{$\emph{partDSA}_{Brier}$}  &     \multicolumn{2}{c}{IPCW}     & \\ \hline
Censoring &Criteria & $1 Fixed$ & $5 Even$ & $5 KM$ & $\emph{partDSA}$ & $CART$ & $L\&C_{NLL}$ \\  \hline 
\hline 
& True Model Size & 2.00 & 2.00 & 2.00 & 2.00 & 3.00 & 3.00 \\ \hline
\multirow{4}{*}{$0\%$/$0\%$}
&Fitted Size&1.642&1.502&1.389&1.456&1.590&1.873\\
&\# Predictors&0.705&0.521&0.496&0.586&0.571&0.838\\
&\# W1-W2&0.621&0.471&0.410&0.491&0.527&0.786\\
&\# W3-W5&0.084&0.050&0.086&0.095&0.044&0.052\\
&$C_p$&0.608&0.604&0.602&0.607&0.603&0.61\\
&$\bar{C}_p$&0.567&0.562&0.559&0.563&0.563&0.571\\
&$L_p$ &0.08&0.086&0.091&0.087&0.083&0.071\\
&$D_p$ &0.618&0.592&0.581&0.6&0.608&0.644\\
\hline
\multirow{4}{*}{$30\%$/$26.3\%$}
&Fitted Size&1.436&1.358&1.285&1.229&1.342&1.573\\
&\# Predictors&0.537&0.448&0.400&0.378&0.332&0.550\\
&\# W1-W2&0.441&0.354&0.315&0.272&0.285&0.516\\
&\# W3-W5&0.096&0.094&0.085&0.106&0.047&0.034\\
&$C_p$&0.603&0.599&0.601&0.594&0.593&0.608\\
&$\bar{C}_p$&0.56&0.556&0.556&0.551&0.552&0.565\\
&$L_p$ &0.076&0.08&0.081&0.084&0.081&0.071\\
&$D_p$ &0.586&0.57&0.571&0.559&0.569&0.606\\
\hline
\multirow{4}{*}{$50\%$/$46.1\%$}
&Fitted Size&1.226&1.256&1.167&1.129&1.170&1.385\\
&\# Predictors&0.291&0.366&0.269&0.186&0.165&0.366\\
&\# W1-W2&0.210&0.274&0.171&0.123&0.122&0.334\\
&\# W3-W5&0.081&0.092&0.098&0.063&0.043&0.032\\
&$C_p$&0.586&0.592&0.59&0.573&0.573&0.605\\
&$\bar{C}_p$&0.547&0.551&0.548&0.538&0.539&0.56\\
&$L_p$ &0.059&0.058&0.06&0.061&0.061&0.055\\
&$D_p$ &0.549&0.56&0.548&0.542&0.545&0.579\\
\hline
\end{tabular}
\end{center}
\label{WebTable:MultiSim1_2nonInf}
\end{table}

\begin{table}
 \caption{Low Signal, Covariate-Independent Setting:
Proportion of the 1000 models at each given size for each method at the three censoring levels.\label{WebTable:MV1_2nonInfModelSize}}
 \begin{center}
 \begin{tabular}{lrrrr}\hline\hline
\multicolumn{1}{l}{}&\multicolumn{1}{c}{Root Node}&\multicolumn{1}{c}{2}&\multicolumn{1}{c}{3}&\multicolumn{1}{c}{4+}\tabularnewline
\hline
{\bfseries 0Cens}&&&&\tabularnewline
~~$L\&C_{NLL}$&$0.458$&$0.287$&$0.207$&$0.048$\tabularnewline
~~$CART_{IPCW}$&$0.559$&$0.320$&$0.096$&$0.025$\tabularnewline
~~$\emph{partDSA}_{IPCW}$&$0.635$&$0.290$&$0.062$&$0.013$\tabularnewline
~~$\emph{partDSA}_{Brier} 1 fixed$&$0.693$&$0.239$&$0.055$&$0.013$\tabularnewline
~~$\emph{partDSA}_{Brier} 5 even$&$0.584$&$0.343$&$0.062$&$0.011$\tabularnewline
~~$\emph{partDSA}_{Brier} 5 KM$&$0.522$&$0.336$&$0.121$&$0.021$\tabularnewline
\hline
{\bfseries 30Cens}&&&&\tabularnewline
~~$L\&C_{NLL}$&$0.620$&$0.228$&$0.120$&$0.032$\tabularnewline
~~$CART_{IPCW}$&$0.727$&$0.215$&$0.048$&$0.010$\tabularnewline
~~$\emph{partDSA}_{IPCW}$&$0.809$&$0.160$&$0.025$&$0.006$\tabularnewline
~~$\emph{partDSA}_{Brier} 1 fixed$&$0.765$&$0.191$&$0.040$&$0.004$\tabularnewline
~~$\emph{partDSA}_{Brier} 5 even$&$0.723$&$0.218$&$0.042$&$0.017$\tabularnewline
~~$\emph{partDSA}_{Brier} 5 KM$&$0.664$&$0.257$&$0.062$&$0.017$\tabularnewline
\hline
{\bfseries 50Cens}&&&&\tabularnewline
~~$L\&C_{NLL}$&$0.729$&$0.184$&$0.071$&$0.016$\tabularnewline
~~$CART_{IPCW}$&$0.854$&$0.127$&$0.015$&$0.004$\tabularnewline
~~$\emph{partDSA}_{IPCW}$&$0.892$&$0.092$&$0.012$&$0.004$\tabularnewline
~~$\emph{partDSA}_{Brier} 1 fixed$&$0.873$&$0.100$&$0.018$&$0.009$\tabularnewline
~~$\emph{partDSA}_{Brier} 5 even$&$0.787$&$0.175$&$0.033$&$0.005$\tabularnewline
~~$\emph{partDSA}_{Brier} 5 KM$&$0.827$&$0.138$&$0.027$&$0.008$\tabularnewline
\hline
\end{tabular}
\end{center}
\end{table}

\processdelayedfloats
\newpage

\section*{Web Appendix D: Descriptive Table for Glioma Dataset}

The prognostic model building ability of the partDSA and CART methods are assessed with the use of
twelve North American Brain Tumor Consortium (NABTC) consecutive Phase II clinical trials for recurrent glioma,  detailed in \citet{Wu01022010}. As indicated in the main document,
the purpose of such trials is to assess the efficacy  of novel therapeutic agents in patients with high-grade gliomas (World Health Organization (WHO) grade III or IV), as glioblastoma patients treated with standard of care have a dismal median survival of 14.6 months \citep{doi:10.1056/NEJMoa043330}.
Below, we expand the analysis of the NABTC trial considered in the main paper, and
evaluate the results of other methods used to define risk groups.

\begin{figure}[!htb]
\begin{center}
\includegraphics[width=6in]{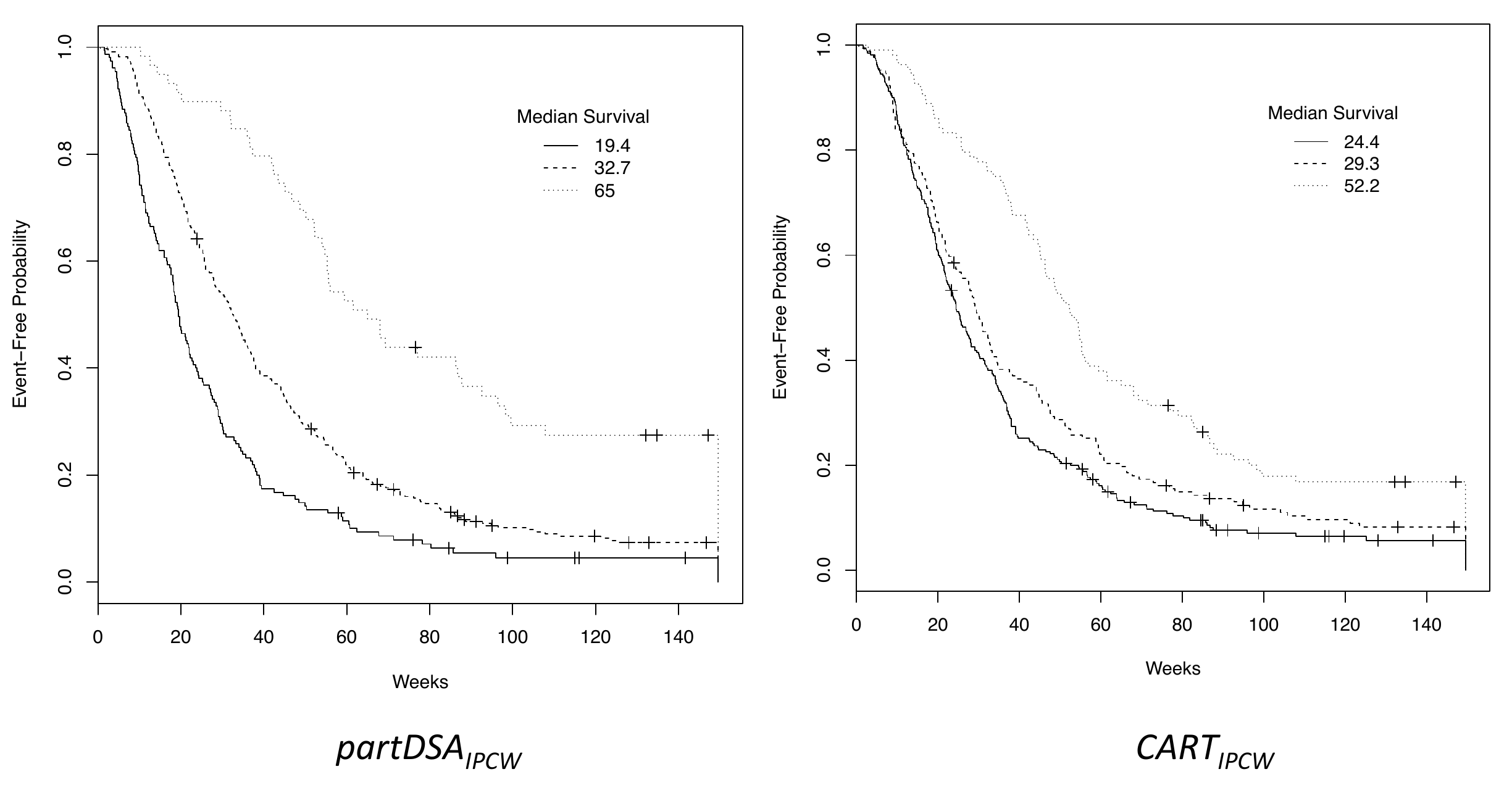}
\caption{{\em Kaplan-Meier Curves for NABTC data analysis in Section
\ref{s:DA}}.} Left panel: $partDSA_{IPCW}$  stratification of patients into three risk groups.
Right panel: $CART_{IPCW}$   stratification of patients into three risk groups.

\label{Fig:KMCurvesSupp}
\end{center}
\end{figure}

The $partDSA_{IPCW}$ method results in three risk groups (defined in Table \ref{Fig:partDSAIPCWData}) and is remarkable as it separates a low risk group with median survival of 65 weeks, the highest of all low risk groups (Fig. \ref{Fig:KMCurvesSupp}, left panel). These 59 patients are $\le 55$, have very high KPS (i.e. $\ge 90$),  no prior TMZ, and if grade IV  were also within 41 weeks of diagnosis.  The intermediate risk group has 335 patients and is defined by prior TMZ, Age, KPS, Grade, time since diagnosis, gender, and baseline anticonvulsant therapy.  The high risk group has 155 patients with median survival of 19.4 weeks and is  defined by prior TMZ,  low KPS ($\le 80$), and gender.
In comparison to the results presented in the main paper, this analysis does not identify
current TMZ or steroids as important predictors.
The $CART_{IPCW}$ tree in Table \ref{Fig:CARTIPCWData} only identifies two variables, prior TMZ and KPS. The 108 patients in the low risk group have a median survival of 52.2 and are defined by a high KPS and no prior TMZ. Interestingly, and similar to the $L\&C_{NLL}$ results shown in the right panel of Fig.~\ref{Fig:KMCurves},  the $CART_{IPCW}$ defined intermediate and high risk groups have a similar survival experience (right panel of Fig.~\ref{Fig:KMCurvesSupp}).

\begin{figure}[!htb]
\label{Fig: KM NABTC Supp}
\begin{center}
\includegraphics[width=6in]{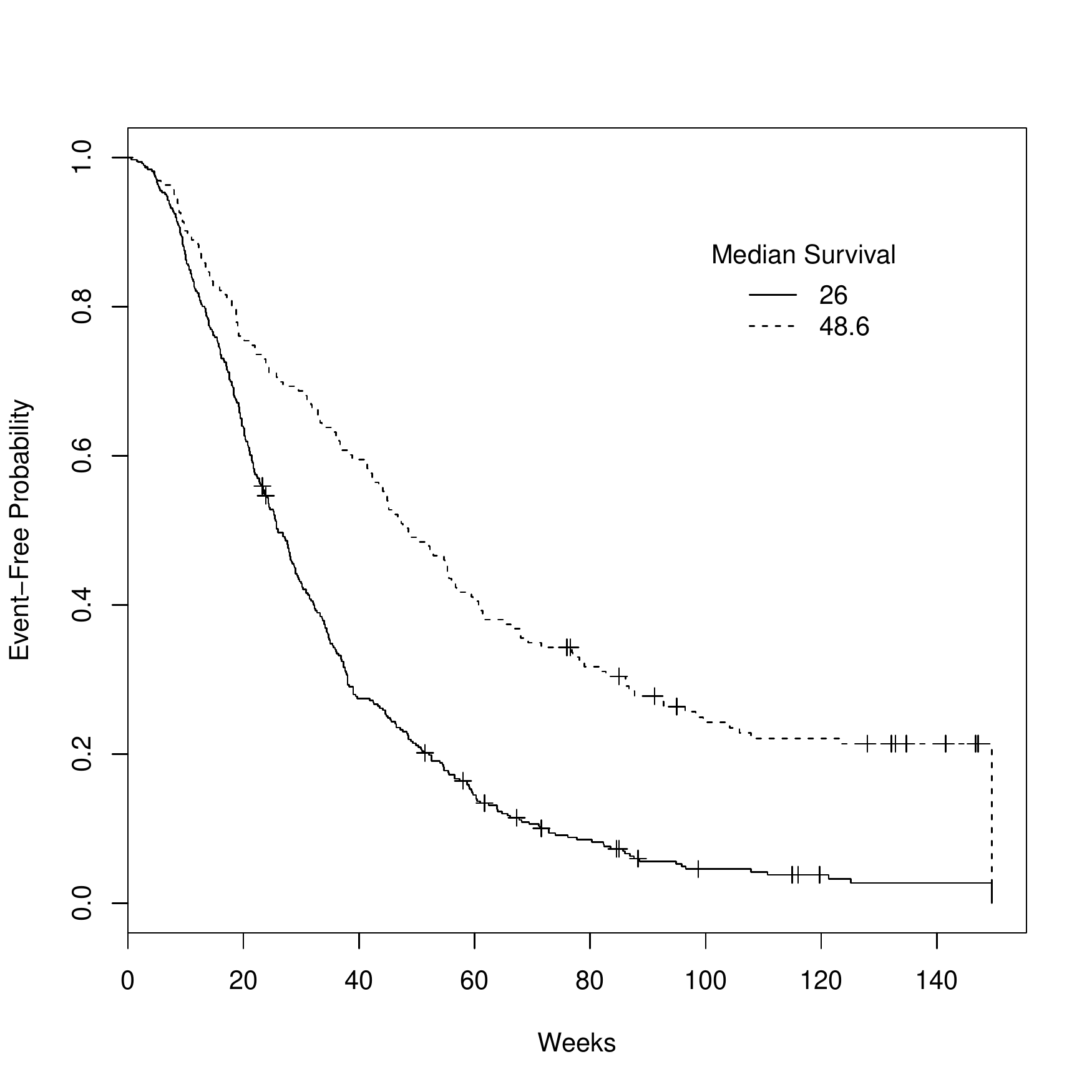}
\caption{{\em Kaplan-Meier Curves for $partDSA_{Brier}(5even)$ analysis of NABTC data in Section
\ref{s:DA}}}

\label{Fig:KMCurvesSupp2}
\end{center}
\end{figure}

The $partDSA_{Brier}(5Even)$ tree only separates out two risk groups (Table~\ref{Fig:partDSABrierEvenData}). The three notable variables in this model are extent of resection (biopsy (Bx) vs. sub-total (ST) vs. gross total (GT)), last known histology (Anaplastic astrocytoma (AA) vs. other), and grade. The 163 patients in the low risk group have median survival of 48.6 weeks, are younger, have $\le 125$ weeks since diagnosis,  and have  typically not had prior TMZ. The survival curves for both groups are shown in Fig.~\ref{Fig:KMCurvesSupp2}. A difference
between this analysis and that based on $partDSA_{Brier}(5KM)$ relates to the time
points used to determine risk groups; in the former, the times tend to be smaller,
and hence the factors identified here may at least partially reflect shorter-term
determinants of survival.

There are several notable differences between the analyses presented here and that in \cite{Wu01022010}. First, we are limited to the 12 NABTC trials whereas \cite{Wu01022010} combined data from 27 NABTC and NCCTG trials. Second, some NABTC trials included both Phase I and Phase II components. For the purposes of this analysis, if a patient was included in both  we only kept the clinical record for the Phase II component so that we had  unique observations. This reduced our sample size from 596 (as reported in \cite{Wu01022010}) to 549. Third, the NABTC trials used KPS instead of ECOG scale (as was used in the NCCTG trials); the results for performance score in
\cite{Wu01022010} reflect both scores, whereas our analyses are limited to KPS only.
Fourth, in \cite{Wu01022010}, current TMZ is used to define an ``adjusted survival
 prediction'' that is subsequently treated in the recursive partitioning
  analysis as an uncensored response variable; in our analyses, current TMZ is
  included as an explanatory variable.

\setlongtables
\renewcommand{\arraystretch}{0.5}
\begin{longtable}{lr} \caption{Baseline demographics for 12 NABTC trials.} \tabularnewline* \hline\hline \label{T:Glioma}
Variable & \multicolumn{1}{c}{NABTC (n = 549; No. [\%])} \tabularnewline* \hline \hline
\endfirsthead
\caption[]{\em (continued)} \tabularnewline*
\hline \hline
\multicolumn{1}{l}{Variable}&\multicolumn{1}{c}{NABTC (n = 549; No. [\%])} \tabularnewline*
\hline \hline
\endhead
\hline
\endfoot
\label{Variable}
{\bfseries Age}&\tabularnewline*
~~Age (years) Median (range)&49 (20-85)\tabularnewline
\hline
{\bfseries Gender}&\tabularnewline*
~~Female&196 (36)\tabularnewline*
~~Male&353 (64)\tabularnewline
\hline
{\bfseries Race/ethnicity}&\tabularnewline*
~~Nonwhite&37 (7)\tabularnewline*
~~White&512 (93)\tabularnewline
\hline
{\bfseries Karnofsky performance score}&\tabularnewline*
~~60&30 (5)\tabularnewline*
~~70&100 (18)\tabularnewline*
~~80&181 (33)\tabularnewline*
~~90&168 (31)\tabularnewline*
~~100&66 (12)\tabularnewline*
~~Missing&4 (1)\tabularnewline
\hline
{\bfseries Year of study entry}&\tabularnewline*
~~1990-1999&174 (32)\tabularnewline*
~~2000-2004&375 (68)\tabularnewline
\hline
{\bfseries Time since initial diagnosis}&\tabularnewline*
~~Time since initial diagnosis (weeks) Mean (range)&97 (10-814) \tabularnewline
\hline
{\bfseries Last known grade}&\tabularnewline*
~~III&147 (27)\tabularnewline*
~~IV&402 (73)\tabularnewline
\hline
{\bfseries Last known histology}&\tabularnewline*
~~Anaplastic astrocytoma&91 (17)\tabularnewline*
~~Anaplastic oligodendroglioma&37 (7)\tabularnewline*
~~Anaplastic oligoastrocytoma&19 (3)\tabularnewline*
~~Glioblastoma multiforme&402 (73)\tabularnewline
\hline
{\bfseries Extent of primary resection}&\tabularnewline*
~~Biopsy&114 (21)\tabularnewline*
~~Subtotal resection&226 (41)\tabularnewline*
~~Gross total resection&145 (26)\tabularnewline*
~~Missing&64 (12)\tabularnewline
\hline
{\bfseries Baseline steroid use}&\tabularnewline*
~~No&244 (44)\tabularnewline*
~~Yes&305 (56)\tabularnewline
\hline
{\bfseries Baseline anticonvulsant use}&\tabularnewline*
~~No&123 (22)\tabularnewline*
~~Yes&426 (78)\tabularnewline
\hline
\newpage
{\bfseries Prior chemotherapy}&\tabularnewline*
~~Missing&1 (0)\tabularnewline*
~~No&117 (21)\tabularnewline*
~~Yes&431 (79)\tabularnewline
\hline
{\bfseries Prior nitrosoureas}&\tabularnewline*
~~No&341 (62)\tabularnewline*
~~Yes&208 (38)\tabularnewline
\hline
{\bfseries Prior TMZ use}&\tabularnewline*
~~No&277 (50)\tabularnewline*
~~Yes&272 (50)\tabularnewline
\hline
{\bfseries Type of treatment center}&\tabularnewline*
~~Academic&549 (100)\tabularnewline
\hline
{\bfseries Number of prior relapses}&\tabularnewline*
~~Missing&60 (11)\tabularnewline*
~~\textless=1&54 (10)\tabularnewline*
~~\textgreater=2&435 (79)\tabularnewline
\hline
{\bfseries Initial low-grade histology}&\tabularnewline*
~~No&436 (79)\tabularnewline*
~~Yes&53 (10)\tabularnewline*
~~Missing&60 (11)\tabularnewline
\hline
{\bfseries Current TMZ}&\tabularnewline*
~~No&367 (67)\tabularnewline*
~~Yes&182 (33)\tabularnewline
\hline
\end{longtable}


	\begin{table}
\centering \caption{$partDSA_{IPCW}$  stratification   of NABTC patients (Sect. \ref{s:DA}) into three risk groups (column 1: low, intermediate (Int), and high). Corresponding median survival in weeks and 95\% confidence intervals (CI) are given in column 2 and number of patients in each group in column 3.  Variables included in the model (columns 4--10) are prior TMZ, age, Karnofsky performance score (KPS), histological grade (Grade), time since diagnosis (Dx), gender, and baseline anticonvulsant use. }
\begin{tabular}{cccccccccc}	
		&			&					& \multicolumn{7}{c}{Variables	}	\\ \cline{4-10}		
Risk	&	Median survival	&	n	&	Prior	&		&		&		&	Time Since	&		&	 Baseline	 \\
 Group	&	 (95\% CI)	&	(549)	&	TMZ	&	Age	&	KPS	&	Grade	&	Dx	&	Gender	&	Anticon	 \\ \hline
\multirow{2}{*}{Low}		&	65 	&	\multirow{2}{*}{59}	&	No	&	$\le 55$	&	$\ge 90$	&	 III	 &		&		&		\\
	&	(54.7, 92.6)	&		&	No	&	$\le 55$	&	$\ge 90$	&	IV	&	$ \le 40.9$	&		 &		 \\ \hline
\multirow{6}{*}{Int}	&	 	&	\multirow{6}{*}{335}	&	Yes	&		&	$\ge 90$	&		&		 &		 &		\\
	&		&		&	No	&		&	$\le 80$	&		&		&	Female	&	No	\\
	&		32.7&		&	No	&	$ > 55$	&	$\ge 90$	&		&		&		&		\\
	&(28.7, 35.9)		&		&	No	&	$\le 55$	&	$\ge 90$	&	IV	&	$ > 40.9$	&		 &		 \\
	&		&		&	Yes	&		&	$\le 80$	&		&	$ \le 38.3 $	&	Female	&		\\
	&		&		&	No	&		&	$\le 80$	&		&		&		&	Yes	\\ \hline
\multirow{3}{*}{High}	&		&	\multirow{3}{*}{155}	&	Yes	&		&	$\le 80$	&		&		 &	 Male	&		\\
	&	19.4	&		&	Yes	&		&	$\le 80$	&		&	$ \ge 38.3$	&	Female	&		\\
	&	 (18, 22.7)	&		&	No	&		&	$\le 80$	&		&		&	Male	&	No	\\  \hline
\end{tabular}
\label{Fig:partDSAIPCWData}
\end{table}

	\begin{table}
\centering \caption{$CART_{IPCW}$  stratification   of NABTC patients (Sect. \ref{s:DA}) into three risk groups (column 1: low, intermediate (Int), and high). Corresponding median survival in weeks and 95\% confidence intervals (CI) are given in column 2 and number of patients in each group in column 3.  Variables included in the model (columns 4--5) are prior TMZ and Karnofsky performance score (KPS). }
\begin{tabular}{ccccc}	
		&			&					& \multicolumn{2}{c}{Variables	}	\\ \cline{4-5}
Risk Group	&	Median survival (95\% CI)	&	n (549)	&	Prior TMZ	&	KPS	\\ \hline

Low	&	52.2 (46.3, 56.9)	&	108	&	No	&	$\ge 90$			\\
Intermediate	&	29.3 (25.3, 33.3)	&	169	&	No	&	$\le 80$			\\
High	&	24.4 (22, 28)	&	272	&	Yes	&				\\ \hline

\end{tabular}
\label{Fig:CARTIPCWData}
\end{table}

	\begin{table}
\centering \caption{$partDSA_{Brier}(5even)$  stratification   of NABTC patients (Sect. \ref{s:DA}) into two risk groups (column 1: low  and high). Corresponding median survival in weeks and 95\% confidence intervals (CI) are given in column 2 and number of patients in each group in column 3.  Variables included in the model (columns 4--10) are Karnofsky performance score (KPS),  age, extent of resection (biopsy (Bx), sub-total (ST), gross total (GT)),  time since diagnosis (Dx), prior TMZ, last known histology (Anaplastic astrocytoma (AA) vs. others), and histological grade (Grade). }
\begin{tabular}{cccccccccc}	
		&			&					& \multicolumn{7}{c}{Variables	}	\\ \cline{4-10}		
Risk 	&	Median survival	&	n	&		&		&	Extent of	&		Time Since		& Prior		 & Last known	&	\\
Group	&	(95\% CI)	&	(549)	&	KPS	&	Age	&	 Resect	&	 Dx	&	 TMZ	&	histology	&	 Grade	 \\ \hline
\multirow{4}{*}{Low}	&		&	163	&		&	$\le 55$	&		&	$\le 126.6$	&	No	&		 &	 III	\\
	&	48.6 &		&		&		&		&	$\le 102.6$	&	Yes	&	AA	&		\\
	&	(42.3, 59.4)		&		&		&	$\le 55$	&	ST/GT	&	$\le 126.6$	&	No	&		 &	IV	 \\
	&		&		&	$\ge 90$	&	$\le 55$	&		&	$\le 126.6$	&	No	&		&		\\ \hline
\multirow{5}{*}{High}	&		&	386	&		&	$>55$	&		&		&	No	&		&		\\
	&	26	&		&		&		&		&		&	Yes	&	not AA	&		\\
	&	(23.7, 29)	&		&		&		&		&	$> 102.6$	&	Yes	&	AA	&		\\
	&	 	&		&	$\le 80$	&	$\le 55$	&		&	$> 126.6$	&	No	&		&		\\
	&		&		&		&	$\le 55$	&	Bx/ST	&	$\le 126.6$	&	No	&		&	IV	\\  \hline
\end{tabular}
\label{Fig:partDSABrierEvenData}
\end{table}

\processdelayedfloats

\processdelayedfloats

\end{document}